\documentstyle[preprint,eqsecnum,aps]{revtex}
\begin{document}
\draft
\title{Separation of Spin and Charge Quantum Numbers in Strongly
Correlated Systems}
\author{Christopher Mudry and Eduardo Fradkin}
\address{Physics Department, University of Illinois at Urbana-Champaign,
1110 West Green Street, Urbana, Illinois 61801-3080}
\date{\today}
\maketitle

\begin{abstract}
In this paper we reexamine the problem of the separation of spin and
charge degrees of freedom in two dimensional strongly correlated
systems. We establish a set of sufficient conditions for the occurence of
spin and charge separation. Specifically, we discuss this issue in the
context of the Heisenberg model for spin-1/2 on a square lattice with nearest
($J_1$) and next-nearest ($J_2$) neighbor antiferromagnetic couplings.
Our formulation makes explicit the existence of a local SU(2) gauge symmetry
once the spin-1/2 operators are replaced by bound states of spinons.
The mean-field theory for the spinons is solved numerically as a function
of the ratio $J_2/J_1$ for the so-called s-RVB Ansatz. A second order
phase transition exists into a novel flux state
for $J_2/J_1>(J_2/J_1)_{{\rm cr}}$. We identify the range
$0<J_2/J_1<(J_2/J_1)_{\rm cr}$ as the s-RVB phase. It is characterized by
the existence of a finite gap to the elementary excitations (spinons)
and the breakdown of all the continuous gauge symmetries. An effective
continuum theory for the spinons and the gauge degrees of freedom is
constructed just below the onset of the flux phase. We argue that this
effective theory is consistent with the deconfinement of the spinons
carrying the fundamental charge of the gauge group. We contrast this result
with the study of the one dimensional quantum antiferromagnet within the
same approach. We show that in the one dimensional model, the spinons of
the gauge picture are always confined and thus cannot be identified with the
gapless spin-1/2 excitations of the quantum antiferromagnet Heisenberg model.
\end{abstract}

\narrowtext

\section{Introduction}
\label{sec:Introduction}

The issue of the possible breakdown of the Fermi-liquid picture
is of considerable interest, both from a practical and conceptual
point of view. The anomalous (with respect to a conventional Fermi-liquid
scheme) properties of the metallic state of high $T_c$ materials
\cite{Bedell 1990}, its proximity to an insulating magnetic state upon doping,
might invalidate the use of Fermi-liquid theory.
The microscopic foundation of Fermi-liquid theory itself
has been a challenging theoretical problem for a long time.
Fermi-liquid theory assumes that there is a one to one relationship
between the low energy excitations of the non-interacting system and the low
energy excitations of the interacting system which preserves quantum numbers
like the spin, charge or particle number.
In contrast, it has been proposed that for certain
{\it strongly interacting} systems,
the low lying excitations are of a
radically different nature than in the non-interacting limit.
For example, the spectrum just above the ground state would be made of
``quasiparticles'' carrying either the spin or the charge quantum numbers
of the non-interacting excitations but not both simultaneously.
This is the scenario of {\it spin and charge separation}
which has been proposed by analogy to the
{\it Hubbard model in one dimension} by some
\cite{Anderson 1987,Anderson 1990}
or to the {\it fractional quantum Hall effect} by others
\cite{Laughlin 1988}.

The purpose of this paper is to reexamine the conditions
under which {\it full} separation of spin and charge occurs
in {\it two dimensional electronic systems}.
It is believed that a necessary condition for this phenomenon
to take place is that the ground state in the spin sector is a
{\it spin liquid} \cite{Anderson 1987,Laughlin 1988,Canright 1990}.
A minimal requirement on a state describing a spin liquid is that it does
not support any long range order breaking the continuuous symmetry under
global spin rotations, for example N\' eel ordering.
However, under these conditions, many different types of
spin liquids can still be constructed.
Of crucial importance to applications to high $T_{\rm c}$ superconductivity
is whether the spin liquid is separated from the excited states
by a gap or not and in the latter case the nature of the
density of states.

The separation of spin and charge is assumed
to occur for a certain range of electronic densities
as a result of the competition between strong electronic correlations
in the form of antiferromagnetic (AF) interactions and the kinetic
energy of the electrons.
We will be interested in the Hubbard model \cite{Hubbard 1964}
with moderate to strong $U$ or its generalization,
the $t-J$ model \cite{tjmodel},
on a two dimensional lattice close to half-filling.
At half-filling, the $t-J$ model
and the $U/t\rightarrow\infty$ limit of the Hubbard model \cite{Emery 1976}
reduce to the antiferromagnetic Heisenberg model for spin-1/2.
The analytical treatment for the separation of spin and charge
usually starts with the factorization  of the creation operator
$c^{\dag}_{i\alpha}$ for the band
electrons occuring in the Hubbard or $t-J$ model
into a pair of operators satisfying fermionic and bosonic
algebra, respectively \cite{slavefermibose,Zou 1988}.
At this point, the assignment for the spin and charge quantum numbers
is arbitrary. We will choose the fermionic operator $s^{\dag}_{i\alpha}$
to carry the spin quantum number ($\alpha=\uparrow,\downarrow$)
but no charge and the state created by $s^{\dag}_{i\alpha}$ is
a {\it bare spinon}. The bosonic operator $h^{\dag}_{i}$
must then carry the charge and the state
it creates is similarly identified with a {\it bare holon}.
Our choice is a matter of convenience. However,
we believe it is well suited for
the description of a spin liquid with the shortest possible
AF correlations, i.e., of the order of the lattice constant.
The alternative option consisting of fermionic holes and bosonic spins
has been used to describe {\it a spin liquid close to the onset of a state
with long range N\' eel ordering}
\cite{Arovas 1988}.

The local phase of the bare spinon and holon operators is arbitrary.
Accordingly, there exists a local gauge U(1) symmetry once the
$c^{\dag}_{i\alpha}=s^{\dag}_{i\alpha}h^{\   }_i$
representation is chosen.
It thus follows that the {\it bare} spinon (holon) operators can not
create physical states.
Indeed, these operators break the local symmetry and
any theory based on a description in terms of spinons and holons
must necessarily involve gauge fields. By construction, all of these theories
are strongly coupled and, thus, the gauge fields vary rapidly.
The likely scenario is then the {\it confinement} of all excitations
transforming non-trivially under the gauge symmetry.
Thus, it would appear that holons and spinons are permanently
confined in bound states and that
there may not be a separation of
spin and charge in terms of spinons and holons carrying gauge quantum numbers.
Nevertheless, several mean-field theories for the bare spinons and
holons have been constructed at half-filling
\cite{Baskaran 1987,Ruckenstein 1987,Kotliar 1988,%
Affleck 1988,Dombre 1989,Wen 1989,Read 1989}
and away from half-filling
\cite{Grilli 1990}.
The challenge to any theory
supposed to describe spin and charge separation is to
construct physical states associated with the spin and charge separately.

In this paper, we will concentrate on the pure spinon sector
at zero-temperature by working
at half-filling in which case the system is a {\it Mott insulator}
\cite{Mottinsulator}.
A special feature of the Heisenberg limit is that
the local U(1) symmetry is enlarged to SU(2)
\cite{Hsu 1988,Dagotto 1988}. We will refer to this symmetry hereafter as the
color symmetry to distinguish it from the symmetry
under global spin rotations.
We also do not have to deal with the complexities
caused by the {\it dynamical interactions} between the spinons and the holons
\cite{Ioffe 1989}. We will
include {\it frustration} through a next-nearest-neighbor antiferromagnetic
interaction in the Heisenberg limit.
The issue of spin and charge separation
then reduces to the existence of spin-1/2 excitations in the
low energy sector of the theory.

Wen \cite{Wen 1991}
has proposed a mean-field theory for the strongly interacting spinons
which preserves the discrete symmetries under
{\it time-reversal} and {\it parity} as well as the space group symmetries of
the lattice. However, for a certain range of values for the ratio of the
competing nearest ($J_1$) and next-nearest ($J_2$) interactions,
his mean-field theory breaks completely the
continuous part of the local color symmetry.
{\it The fundamental assumption} is that an
Anderson-Higgs mechanism
\cite{SSB1},
driven purely by the quartic spinon interaction,
takes place {\it and} that its occurrence is sufficient to
render the mean-field prediction of the existence of spinons in the physical
spectrum robust to the strong gauge fluctuations caused by the
constrained nature of the system.
This BCS-like ground state is highly unconventional.
Pairing takes place on the direct lattice and the
``spontaneously broken'' local gauge symmetry is not U(1) but SU(2).
{\it It is thus  argued that through the macroscopic condensation
of spinon bound states (the typical size of which is the lattice spacing),
the physical spectrum develops a gap to all excitations
among which some carry the spinon quantum number}.
Frustration is the trigger for this mechanism.
A related mechanism has also been proposed by Read and Sachdev
in their analysis of the Sp(N) $J_1-J_2-J_3$ antiferromagnetic
Heisenberg model \cite{Sachdev 1991}.

However, it is not at all clear to us that the spinon/holon picture is
the appropriate one for the description of spin and charge separation.
In fact, to our knowledge, it has never been demonstrated that
the separation of spin and charge known to occur in many one dimensional
electronic models belonging to the Luttinger liquids universality class
\cite{Haldane 1981},
becomes transparent in the spinon/holon representation.
A first example is the Hubbard model on a linear chain.
There, the separate spin-1/2 and charge sectors are
created by soliton-like operators \cite{Emery 1979} which,
when expressed in terms of the band electron operator
$c^{\dagger}_{i\alpha}$, are highly non-local
and therefore cannot be described in any simple way by the spinon or holons.
Another example is the antiferromagnetic quantum spin-1/2 Heisenberg chain
with nearest-neighbor interactions for which the lowest excitations above
the ground state are known to transform like the fundamental
representation of SU(2) under spin rotations \cite{Faddeev 1981}.
However, these excitations are again created by soliton-like operators
which are related to the original spin-1/2 operators through a highly
non-linear and non-local mapping, the Jordan-Wigner transformation
\cite{Jordan 1928,Luther 1975,Haldane2 1981}.
Nevertheless, the spinon/holon picture offers one of
the few analytical methods to study the issue of separation of spin and charge
provided the mean-field predictions are not taken at face value.
Moreover, aside from this issue,
it offers a powerful technique to study non-Fermi liquid behaviour,
once the local gauge invariance is properly accounted for
\cite{Lee 1992}.

In this paper, we will argue that
the mechanism thought by Wen can but does not necessarily
insure that the spinons remain in the physical spectrum.
To this end, we have constructed an effective field theory for the low lying
spinon modes in the presence of smooth fluctuations of the gauge modes.
This effective field theory describes the low energy physics close to
a second order phase transition from Wen's s-RVB phase
for $0<J_2/J_1<(J_2/J_1)_{\rm cr}$ into a gapless phase that,
we found, is present at the mean-field level.
The gapless state for $J_2/J_1>(J_2/J_1)_{\rm cr}$ is qualitatively
analogous to the flux state of Affleck and Marston \cite{Marston 1989}.
However, in our problem, the gapless phase occurs at the mean-field level
as a result of the competition between nearest ($J_1$) and next-nearest
($J_2$) neighbors. Hereafter, we refer to the gapless state as the
flux state even though it is not the same state as in
\cite{Marston 1989}. Close to and in the flux phase,
the low energy spinon modes and large scale lattice gauge fluctuations
are equivalent to a relativistic theory in 2+1 space-time of four Dirac
fermions minimally coupled to non-Abelian gauge fields.
The multiplication of the fermion species (flavors)
is the result of a degeneracy of the mean-field spectrum. Each
flavor can be thought of as representing low lying fermionic excitations
around one of the isolated but degenerate minima of the Brillouin Zone.
The gauge fields are long wavelength fluctuations of the soft modes
residing on the next-nearest-neighbor links of the lattice,
i.e., the SU(2) color degrees of freedom.
The nearest-neighbor exchange coupling $J_1$ causes the
different fermionic flavors to interact through effective scalar fields.
These scalar fields represent long wavelength fluctuations
of the nearest-neighbor link degrees of freedom.
It is important to stress that these bosonic degrees of freedom
have no existence of their own independently of the spinons.
{\it They represent collective modes for the spinons}.
Just below the onset of the flux phase, the scalar gauge fields acquire
expectation values which break all continuous gauge symmetries down
to a discrete subgroup. Integration of the spinons then yields a local
effective theory for the non-Abelian gauge fields and the scalar fields.
We give {\it a non-perturbative} argument for the existence of the color
quantum number of the spinon in the spectrum of the bosonized effective
theory, provided both kinetic energy scales for the gauge fields and
scalar fields (Higgs) are large enough.
The calculation of these two scales will be published elsewhere
\cite{Mudry}.
This mechanism is different from the one constructed from the
chiral spin liquid since it does not rely on the breaking
of parity and time reversal, which at the level of the effective continuum
action implies the presence of a Chern-Simon term \cite{Wen 1989}.
Indeed, although, the four Dirac fermions acquire a mass it comes with
an alternating sign for each flavor, and thus, upon fermionic
integration, no net Chern-Simon term results.

In section \ref{sec:equiv},
we rapidly describe how to map the Heisenberg model into
a SU(2) lattice gauge theory.
Our strategy to study the issue of spin and charge separation is
outlined in section \ref{sec:Deconfinement}.
In section \ref{sec:s-RVBMF}, the numerical solutions of
the saddle-point equations for the s-RVB Ansatz proposed by Wen are
presented. A generalization of the s-RVB Ansatz to higher dimensions
is also included.
A field theory for the low lying excitations in the vicinity
of the flux phase is derived in section \ref{sec:fluc}.  We contrast
our results with an application of our formalism to
the one dimensional Heisenberg chain in section \ref{sec:onedim}.
Our conclusions are presented in section \ref{sec:concl}.


\section{The equivalence of the Heisenberg model for spin-1/2
to a SU(2) lattice gauge theory}
\label{sec:equiv}

The band electron operator $c_{\uparrow(\downarrow)}$
of the Hubbard model can always be recast in the form
\cite {Hubbard 1965,Zou 1988}
\begin{eqnarray}
c_{\uparrow(\downarrow)}=&&\;
|\;0\;><\;\uparrow(\downarrow)\;|\; +(-)\;
|\;\downarrow(\uparrow)\;><\;\uparrow\downarrow\;|
\\
=&&\;
h^{\dag}\; s^{\   }_{\uparrow(\downarrow)}\; +(-)\;
s^{\dag}_{\downarrow(\uparrow)}\; d^{\   },
\nonumber
\end{eqnarray}
provided the creation operators $h^{\dag}$, $s^{\dag}_{\sigma}$
and $d^{\dag}$ for the holon, spinon and doublon states, respectively,
satisfy the appropriate commutation relations.
{\it In the limit of strong on site repulsion, the term involving the doublon
state is neglected} \cite {Zou 1988}.
We will choose the spinon operator to obey fermionic
commutation relations.

At half-filling, the Hubbard model reduces to the Heisenberg
model for spin-1/2
\cite{Emery 1976}:
\begin{equation}
H=\sum_{<ij>}\; J_{ij}\; \vec S_i\cdot\vec S_j
\label{Heisenberghamiltonian}
\end{equation}
where $<ij>$ is an ordered pair of sites on an arbitrary lattice $\Lambda$,
$J_{ij}$ are antiferromagnetic coupling constants and
\begin{equation}
\vec S_i=
{1\over2}\;
s^{\dag}_{i\alpha}\;
\vec\sigma^{\   }_{\alpha\beta}\;
s^{\   }_{i\beta },
\label{spininspinonbasis}
\end{equation}
$\vec\sigma$ being the Pauli matrices.
The spinon representation, Eq.\ (\ref{spininspinonbasis}), for the spin-1/2
degrees of freedom must be supplemented with {\it any} of the three constraints
\begin{equation}
\openone=
s^{\dag}_{i\alpha}\;
\delta^{\   }_{\alpha\beta}\;
s^{\   }_{i\beta }\;
\Leftrightarrow\;
0\;=\; s^{\dag}_{i\uparrow}\; s^{\dag}_{i\downarrow}\;
\Leftrightarrow\;
0\;=\; s^{\   }_{i\downarrow}\; s^{\   }_{i\uparrow}
\label{singleoccupancy}
\end{equation}
for all sites $i$ of the lattice.

{}From the fully symmetric tensor $\delta_{\alpha\beta}$
and the fully antisymmetric tensor
$\epsilon_{\alpha\beta}$
of SU(2), the two bilinear forms
\begin{equation}
\chi^{\dag}_{ij}\; =\;
s^{\dag}_{i\alpha}\;
\delta^{\   }_{\alpha\beta}\;
s^{\   }_{j\beta }
\end{equation}
and
\begin{equation}
\eta^{\dag}_{ij}\; =\;
s^{\dag}_{i\alpha}\;
\epsilon^{\   }_{\alpha\beta}\;
s^{\dag}_{j\beta },
\end{equation}
can be used to describe a singlet pairing of the two spin-1/2
located on site $i$ and $j$, respectively
\cite{Affleck 1988,Baskaran 1987}.
Indeed, the identity
\begin{equation}
\vec S_i\cdot\vec S_j\; =\;
-{1\over4}\; \eta^{\dag}_{ij}\; \eta^{\   }_{ij}\; -\;
 {1\over4}\; \chi^{\dag}_{ij}\; \chi^{\   }_{ij}\; +\;
 {1\over4}\; \openone
\end{equation}
holds in the Hilbert space of one spinon per site.
A {\it spin liquid} which, by definition,
should not show any long range magnetic order, implies,
in the spinon picture, the exponential decay with separation $|i-j|$
of the vacuum expectation values $<\eta^{\dag}_{ij}>$ or $<\chi^{\dag}_{ij}>$.

The dynamics of these bilinear forms can be obtained from the vacuum
persistence amplitude
\begin{eqnarray}
Z\;=&&
\;
\int{\cal D}\;\left[\vec a_{\hat{\rm o}}\right]\;
\int{\cal D}\;\left[ s^{*}\right]
{\cal D}\;\left[s^{\ }\right]\;
e^{+{\rm i}\int dt\; L'}
\end{eqnarray}
where the lattice Lagrangian is
\begin{eqnarray}
L'\;=&&
\;\sum_i\; s^{*}_{it\alpha}\;{\rm i}\partial_t\; s^{\ }_{it\alpha}
\nonumber\\
-&&\;\sum_i\;
\left(
{1\over2}\; a^-_{\hat{\rm o}it}\; \eta^{* }_{iit}+
{1\over2}\; a^+_{\hat{\rm o}it}\; \eta^{\ }_{iit}+
{1\over2}\; a^3_{\hat{\rm o}it}\; (\chi^{*}_{iit}-1)
\right)
\nonumber\\
+&&\;\sum_{<ij>}\;
{J_{ij}\over4}\;
\left(
\eta^{*}_{ijt}\eta^{\ }_{ijt}
\;+\;
\chi^{*}_{ijt}\chi^{\ }_{ijt}
\right).
\label{firstlatticelagrangian}
\end{eqnarray}
The integration over the Lagrange multipliers
\begin{equation}
\pmatrix
{
a^-_{\hat{\rm o}it}\cr
\hfil    \cr
a^+_{\hat{\rm o}it}\cr
\hfil    \cr
a^3_{\hat{\rm o}it}\cr
}\;
=\;
\pmatrix
{
{1\over2}\;(a^1_{\hat{\rm o}it}\;-\;{\rm i}\ a^2_{\hat{\rm o}it})\cr
\hfil    \cr
{1\over2}\;(a^1_{\hat{\rm o}it}\;+\;{\rm i}\ a^2_{\hat{\rm o}it})\cr
\hfil    \cr
a^3_{\hat{\rm o}it}\cr
}
\end{equation}
enforces the constraint of single occupancy on the spinon Hilbert space
in a redundant way. But this redundancy allows for the mapping of
Eq.\ (\ref{firstlatticelagrangian})
into the Lagrangian of a SU(2) lattice gauge theory
\cite{Hsu 1988}, if
a Hubbard-Stratanovich transformation with respect to the composite fields
$\eta$ and $\chi$ is performed first.

With a particle-hole transformation of the spinons,
our final Lagrangian will then take the form \cite{Dagotto 1988}
\begin{eqnarray}
L\; =&&\;
\sum_i\;
\psi^{* }_{it}\;
\left(\;
{\rm i}\partial_t-A_{\hat{\rm o}it}\;
\right)\;
\psi^{\ }_{it}
\label{finallatticelagrangian}\\
-&&\;
\sum_{<ij>}{J_{ij}\over4}
\left(
|{\rm det}\; W^{\   }_{ijt}|+
(
\psi^{* }_{it}W^{\   }_{ijt}\psi^{\ }_{jt}+
\psi^{* }_{jt}W^{\dag}_{ijt}\psi^{\ }_{it}
)
\right).\nonumber
\end{eqnarray}
Here, the $\psi$'s are \cite{particle-hole}
\begin{equation}
\psi_{it}=
\pmatrix
{
s^{\   }_{it\uparrow  }\cr
s^{*   }_{it\downarrow}\cr
}.
\label{particleholetrs}
\end{equation}
The $A_{\hat{\rm o}}$'s
belong to the fundamental representation of the su(2) Lie algebra
\begin{equation}
A_{\hat{\rm o}it}\;=\; {1\over2}\; \vec a_{\hat{\rm o}it}\; \cdot\; \vec\sigma.
\end{equation}
Finally, the $W$'s are 2$\times$2 matrices of the form
\begin{equation}
W^{\   }_{ijt}=
\pmatrix
{
-X^{\ }_{ijt}&-E^{\ }_{ijt}\cr
-E^{* }_{ijt}&+X^{* }_{ijt}\cr
},
\end{equation}
which satisfy
\begin{equation}
W^{\   }_{ijt}\;=\;W^{\dag}_{jit}.
\end{equation}
The entries $E$ and $X$ of the $W$'s are the Hubbard-Stratanovich degrees
of freedom associated with the spinon bilinears $\eta$ and $\chi$,
respectively.

The lattice Lagrangian in
Eq.\ (\ref{finallatticelagrangian})
is left unchanged by the local
gauge transformations
\begin{eqnarray}
&&
\psi^{\   }_{it}\;\rightarrow\;
\psi'      _{it}\;=\; U^{\   }_{it}\; \psi^{\   }_{it},
\nonumber\\
&&
A_{\hat{\rm o}it}\;\rightarrow\;
A_{\hat{\rm o}it}'\;=\;
U^{\   }_{it}\;
A_{\hat{\rm o}it}\;
U^{\dag}_{it}\;+\;
\left(
{\rm i}\partial_t\;
U^{\   }_{it}
\right)
U^{\dag}_{it},
\nonumber\\
&&
W^{\   }_{ijt}\;\rightarrow\;
W'      _{ijt}\;=\;
U^{\   }_{it}\;
W^{\   }_{ijt}\;
U^{\dag}_{jt},
\label{latticegaugeinv}
\end{eqnarray}
for all $U^{\   }_{it}\ \in$ SU(2).
This local symmetry will be called a {\it color symmetry}. It is a different
symmetry from the one generated by global spin rotations. Indeed, under
the particle-hole transformation Eq.\ (\ref{particleholetrs}),
the spin-1/2 operators of Eq.\ (\ref{spininspinonbasis})
are mapped into
\begin{eqnarray}
&&S^1_i\;=\;
+{1\over2}\;
\left(\;
\psi^{\dag}_{i1}\;\psi^{\dag}_{i2}\;+\;
\psi^{\   }_{i2}\;\psi^{\   }_{i1}\;
\right)\;\equiv \;
+{1\over2}\;
\left(\;
b^{\dag}_i+b^{\   }_i
\right),
\nonumber\\
&&S^2_i\;=\;
-{{\rm i}\over2}\;
\left(\;
\psi^{\dag}_{i1}\;\psi^{\dag}_{i2}\;-\;
\psi^{\   }_{i2}\;\psi^{\   }_{i1}\;
\right)\;\equiv \;
-{{\rm i}\over2}\;
\left(\;
b^{\dag}_i-b^{\   }_i
\right),
\nonumber\\
&&S^3_i\;=\;
+{1\over2}\;
\left(\;
\psi^{\dag}_{i}\;\psi^{\   }_{i}\;-\;1
\right)\;\equiv \;
+{1\over2}\;
\left(\;
m^{\   }_i-1
\right).
\label{spininpsibasis}
\end{eqnarray}

The bilinears $b$ and $m$ defined above are left unchanged by the local
color transformation Eq.\ (\ref{latticegaugeinv}) and thus the Heisenberg
interaction explicitly transforms like a color singlet
when expressed in terms of the $\psi$'s.
Spin-spin correlations can be obtained from the generating functional
\begin{eqnarray}
Z[\ \vec J\ ]\;=&&\;
\int{\cal D}\ \left[W^{\dag}\right]{\cal D}\ \left[W\right]
\int{\cal D}\ \left[\vec a_{\hat{\rm o}}\right]
\int{\cal D}\ \left[\psi^{* }\right]{\cal D}\left[\psi^{\ }\right]\;
\nonumber\\
\times&&
e^
{+{\rm i}\int dt
\left(\
L\ +\ \sum_i\ \vec J_i\cdot \vec S_i\
\right)
}
\end{eqnarray}
where the source term is written in terms of
the bilinears $b$ and $m$ defined by
Eq.\ (\ref{spininpsibasis}).

It is important to realize that the gauge degrees of freedom in
Eq. (\ref{finallatticelagrangian})
are not independent from the fermionic degrees of freedom.
For example, neither $A_{\hat{\rm o}}$ nor the SU(2) factor of $W_{ij}$
possess a lattice version of the kinetic energy. Their presence is simply
a device to project the Fock space ${\cal F}$
which is generated cyclically from the
vacuum state $|0>_{\psi}$ defined by
$$
\psi_{ia}\; |0>_{\psi}\;=\;0,\quad \forall i\;\in\;\Lambda,\quad a\;=\;1,2,
$$
onto the physical Hilbert space which is the tensorial product over
all sites $i$ of the vector spaces spanned by the color singlet states
$|0>_{\psi_i}$ and $b^{\dag}_i|0>_{\psi_i}$
(see \cite{particle-hole}).

\section{Spin liquid, deconfinement and spin and charge separation.}
\label{sec:Deconfinement}

In the previous section we have made use of the
mapping of the spin-1/2
antiferromagnetic Heisenberg model into a SU(2) lattice gauge theory
with the hope that the latter formulation is better suited for the description
of a spin-liquid ground state and the possible existence of spin-1/2
excitations. The fact that the problem is, in effect, a gauge
theory indicates that in all likelyhood the spinons will necessarily be
permanently confined inside SU(2) color singlet bound states. This is so
since, below four space-time dimensions all (compact) gauge theories are
generally in a confined phase. If this is
the case, there is little hope that this mapping may be of great help in
producing a mechanism for the separation of spin and charge quantum
numbers in the physical spectrum of the system. Hence, the problem that
needs to be solved is to find mechanisms that separate the spin
and charge {\it either} in a manner consistent with the confinement of spinons
{\it or} by dynamically driving the system into a deconfined state.
In this section we will show that Wen's s-RVB state effectively realizes
the second scenario. In section \ref{sec:onedim}, we show that the
first scenario is realized in the one dimensional version of Wen's state.

In order to describe a spin liquid,
an eigenstate of the Hamiltonian of Eq.\ (\ref{Heisenberghamiltonian})
should show exponential decay with separation for all spin-spin
correlations which break the global rotational spin symmetry.
For example, to avoid N\' eel ordering
we can imagine partitioning the lattice into distinct ordered pairs
and choosing for each such pair the singlet state of the four dimensional
Hilbert space associated with the two spin-1/2 degrees of freedom.
The expectation value of $\vec S_i\cdot\vec S_j$ in this state is only
non-vanishing when $i$ and $j$ belong to the same ordered pair. However,
such a state, the so-called valence bond state \cite{Kivelson 1987},
breaks the lattice symmetry.
If we require ``featurelessness'' from a spin liquid, then
an easy remedy for a valence bond state with nearest-neighbor pairing,
say on a square lattice, is to demand a uniform expectation value of the
valence bond ordering operators for any spin and its four nearest-neighbors.
This leads us naturally to the RVB overcomplete basis
\cite{Anderson 1987} for Hamiltonian
Eq.\ (\ref{Heisenberghamiltonian}).
However, the RVB basis is extremely difficult to work with.

On the other hand, it appears to be very easy to implement
{\it featurelessness} and
{\it exponential decay of the spin-spin correlations}
in the gauge picture, Eq.\ (\ref{finallatticelagrangian}).
Featurelessness amounts to selecting only those gauge equivalent
classes of configurations
${\cal C}_{\{A_{\hat{\rm o}},\; W\}}$
which are closed under the action of the lattice point group,
i.e., gauge configurations on which the action of a point group
transformation is equivalent to a gauge transformation.
Exponential decay of the spin-spin correlations requires
that within the class ${\cal C}_{\{A_{\hat{\rm o}},\; W\}}$
\begin{equation}
|{\rm det}\;W_{ij}|\;\propto\; e^{-|i-j|}
\end{equation}
is satisfied, although this condition might not be sufficient
(overcompleteness of the RVB basis).

This suggests the following
{\it mean-field approximation for a spin liquid}:

\noindent$(i)$ Choose a gauge field configuration consistent with
the requirements of a spin liquid.

\noindent$(ii)$ Solve the fermionic sector in the background  of the
gauge fields given in step (i).

\noindent$(iii)$ Impose the self-consistency condition that the background
gauge fields given in step (i) are the expectation value of the appropriate
fermionic bilinear forms in the ground state found in step (ii).

The drawback of this approach is that the
self-consistency condition only satisfies the local constraint
of single occupancy for the spinons, Eq.\ (\ref{singleoccupancy}),
on average.
In the worst case scenario, one could imagine that all the eigenstates
of the mean-field Hamiltonian for the spinons belong to the unphysical
Hilbert space, although on average the constraint of single occupancy is
satisfied. For example, at half-filling, a state with all but two sites singly
occupied by spinons is orthogonal to the physical subspace of the fermionic
Hilbert space.
In the language of the $\psi$'s, single occupancy on any given site
is the local constraint of Gauss's law required by the local gauge
invariance, namely the local physical states are the color singlets
$|0>_{\psi}$ or $b^{\dag}_i\;|0>_{\psi}$.
It is then easy to convince oneself that the ground state and the
single particle excitations of the mean-field theory
have always a non-vanishing projection onto the unphysical Hilbert space.
Indeed, the mean-field ground state is a Slater determinant for the $\psi$'s
which can be obtained by a Fourier transformation of
\begin{equation}
\prod_{k\in\Omega_0}\psi^{\dag}_{k1}\psi^{\dag}_{k2}\;|0>_{\psi},
\end{equation}
where $\Omega_0$ denotes the set af all $|\Lambda|/2$ wave vectors entering
in the Slater determinant of the mean-field ground state. The mean-field
ground state and the excited states are thus convolutions in direct space
whereas the Fourier transformation of the gauge invariant
operator $b^{\dag}_i=\psi^{\dag}_{i1}\psi^{\dag}_{i2}$ is a convolution
in reciprocal space.

Since the problem of the quantum numbers of the physical states is
non-perturbative by nature, it is important to take into account
the quantum fluctuations of the gauge fields around the mean-field.
The problem we end up with is that of fermions which are strongly
coupled with gauge fields. If the mean-field theory predicts a gap
in the fermionic spectrum a simpler picture can be drawn. In this
case it is possible to integrate out the fermions to obtain a local theory.
Although the resulting effective theory will be constructed
from purely bosonic dynamical fields
(which can be viewed as local fermionic composites), its spectrum can contain
states with fermionic quantum numbers.
We will show below that this is
the case for the s-RVB Ansatz where the effective action involves gauge
fields locally coupled to bosonic matter (scalars) which represent
bound states of the spinons. We will also see that these bound states
have simple transformation properties under the symmetries of the system.
Of particular importance is that we will be able to find bound states
that transform non-trivially under the local SU(2) invariance. Hence, at
least qualitatively, our problem is closely related to that of finding
the nature of the ground state of scalar fields coupled to SU(2)
gauge fields. This is a problem that has been investigated quite
extensively in the context of Lattice Gauge Theory.
In what follows we will review those results,
make connections with our problem and then draw conclusions. It will be
apparent from our discussion that the key to the mechanism is to
determine the transformation properties of the bound states.

Let us set the notation and define a set of scalar fields $\Phi_i$ on the
sites $i$ of a $d+1$ dimensional space-time hypercubic lattice
$\Lambda$ and a set of SU(2) gauge fields
$V_{ij}$ on the ordered links $<ij>$ of the same lattice.
The gauge fields transform like group elements whereas the matter
fields will transform under some representation of SU(2).
In practice we will consider two cases: a)
matter in the {\it fundamental} (or {\it spinor}) representation of SU(2)
color and b) matter in the {\it adjoint} (or {\it vector}) representation of
SU(2) color. A simple but generic action for this system is
\begin{eqnarray}
S\;=&&\;
\beta\;
\sum_{<ijkl>}\;
\left(\;
1\;-\;
{1\over4}\;{\rm tr}\;
\left[\;
V^{\   }_{ij}\;
V^{\   }_{jk}\;
V^{\dag}_{lk}\;
V^{\dag}_{il}\; +\;{\rm H.C.}\;
\right]
\right)
\nonumber\\
-&&
\kappa\;\sum_{<ij>}\;
\left(\;
(
\Phi^{\   }_i,
V^{\   }_{ij}
\Phi^{\   }_j
)\;
+\;{\rm H.C.}
\right)
\nonumber\\
+&&\lambda\sum_i
\left(
\left(
\Phi_i,
\Phi_i
\right)-1
\right)^2
+
\sum_i
\left(
\Phi_i,
\Phi_i
\right),
\nonumber\\
\beta\geq&&0,\quad\kappa\geq0,\quad\lambda>0.
\label{fradkinaction}
\end{eqnarray}
The first term of Eq. (\ref{fradkinaction}), which is a sum over oriented
plaquettes $<ijkl>$, represents the gauge invariant kinetic energy for
the SU(2) gauge fields while the rest of the
action describes the dynamics of the matter and their coupling to the
gauge fields. The scalar product of the matter fields $\Phi$
in a given representation is denoted by $(\cdot,\cdot)$.
Finally, we have also allowed for a fluctuation of the norm
of the matter field which, however, is irrelevant to the problem of
confinement.

The properties of the phase diagram in the $\beta-\kappa-\lambda$ space for
Eq.\ (\ref{fradkinaction})
which are important to us depend only on the dimensionality of space-time
and on the transformation properties of the matter fields
\cite{Fradkin 1979,Evertz 1988}.
We will distinguish two situations. In one case, the  matter field
$\Phi$
is in the fundamental representation, in the other it is in
the adjoint. Those two representations differ in one crucial aspect.
The fundamental representation is one to one whereas the adjoint is two to
one since it fails to distinguish between the two elements of the center
${\rm C_{SU(2)}}$ of the group SU(2) color
\cite{center}.

We first start our exploration of the phase diagram from the point
$(\beta,\kappa,\lambda)=(0,0,\infty)$. By letting
$\lambda\rightarrow\infty$, we have frozen the {\it norm} of the
matter field
to the value  one. When we vary $\beta$ from 0 to $\infty$, we obtain a pure,
non-Abelian lattice gauge theory. Along this line, the
Wilson loop $W_{\ {\rm f(a)}}^{\Gamma}$ for sources
in the fundamental (adjoint) representation
\begin{equation}
W_{\ {\rm f(a)}}^{\Gamma}\;=\;
<\;
{\rm tr}\; \prod_{<ij>\in\Gamma}\; V_{ij}
\;>
\end{equation}
where $\Gamma$ is a closed path on the hypercubic lattice,
can be used to detect a phase transition.
For an appropriate choice of $\Gamma$,
\begin{equation}
\Gamma=\Box_R^T,
\end{equation}
the potential
\begin{equation}
V_{{\rm f(a)}}(R)\equiv -\lim_{T\rightarrow\infty}
{1\over T}\ln W_{\ {\rm f(a)}}(T,R)
\end{equation}
measures the energy of two static
(infinitely heavy) sources
in the fundamental (adjoint) representation
which are separated by a distance $R$.

For small values of $\beta$, i.e., strong gauge coupling, the static
potential is linear (accordingly the Wilson loop obeys an {\it area law})
and the {\it static sources are confined}.
The lower critical space-time dimension is 4 for non-Abelian groups.
We are interested in  space-time of dimension 2+1 and therefore do
not encounter a phase transition suggesting that for larger values of
$\beta$ deconfinement (the Wilson loop obeys a {\it perimeter law})
of the static sources occurs \cite{Wilsonloop}.

At the point $(\infty,0,\infty)$, the pure gauge
theory freezes into any of the classical gauge configurations
with all plaquette variables set to unity. The simplest choice is
to have all $V_{ij}$ set to unity. We then start moving along the line
$(\infty,\kappa,\infty)$. The action
for the matter field in the fundamental (adjoint) representation and
with frozen norm
is equivalent to that of a classical O(4) (O(3)) Heisenberg model.
For space-time dimension larger than 2 and irrespectively of the matter
field
representation, a second order phase transition will occur  when $\kappa$
is larger than $\kappa^{\rm f(a)}_{\rm cr}$
\cite{O(n)model}.

As of now, the choice for the representation of the matter field had no
impact on the existence of a phase transition. However, along the line
$(\beta,\infty,\infty)$, the physics will differ
dramatically. At $(\infty,\infty,\infty)$,
the matter field in the fundamental (adjoint) representation
are frozen into a classical configuration, say $\Phi_0$,
which minimizes the action and breaks the global O(4) (O(3)) symmetry,
provided $d+1>2$. By decreasing the value of $\beta$, the gauge fields
are unfrozen. However, the matter field configuration $\Phi_0$ acts
like an
external symmetry breaking magnetic field through the hopping term. Indeed,
under a gauge transformation
\begin{equation}
(\Phi_0,V_{ij}\Phi_0)\rightarrow
(\Phi_0,U_iV_{ij}U_j^{-1}\Phi_0).
\end{equation}

The adjoint representation differs from the fundamental representation in
that if one chooses to multiply $V_{ij}$ by the representative of $-\sigma^0$
on the left and $+\sigma^0$ on the right, then the hopping term remains
unchanged in the adjoint while it changes sign in the fundamental.
In other words, there will always be a {\it residual  local} ${\cal Z}_2$
symmetry along $(\beta,\infty,\infty)$ in the adjoint representation.
Hence, if the matter field breaks all the continuous symmetry, no phase
transition
takes place in the fundamental representation, while a first order
phase transition takes place at $\beta^{\rm a}_{\rm c}$
in the adjoint representation \cite{Wegner 1971}.

Finally, along the line $(0,\kappa,\infty)$, the action describes a
theory of non-interacting links (see \cite{Drouffe 1983}) which does not
undergo a phase transition.

We can extract valuable informations from
Eq. (\ref{fradkinaction}) in the limiting cases where either the matter
field
or the gauge degrees of freedom  are passive bystanders to the dynamics.
In the interior of the phase diagram of  Fig.~\ref{phasediagram},
gauge and matter degrees of freedom do not decouple from each other.
Consequently, the Wilson loop for static charges carrying the same
representation as the
matter field {\it always} obeys the perimeter law \cite{Evertz 1987}.
Indeed, pair creation of matter,
which causes the screening of the static color charges \cite{Kogut 1974},
are induced by quantum fluctuations.

It also follows from this argument
that for  the matter in the fundamental representation,
all the Wilson loops, irrespectively of the representation of the
static sources, will obey the perimeter law in the interior of
the phase diagram \cite{PLforallWL}.
Wilson loops cannot then be used anymore as order parameters.
In fact, it was proven
\cite{Osterwalder 1978,Fradkin 1979}
that there exists a strip of analyticity joining the small $\beta$
and small $\kappa$ region ({\it Confinement region})
to the large $\beta$
and large $\kappa$ region ({\it Higgs region})
of the phase diagram,
in which all the gauge invariant Green functions are analytic.
In short, there is only one phase, the so-called {\it
Confinement-Higgs
phase}, in the phase diagram $(\beta,\kappa,\infty)$ when the matter
is in the fundamental representation. A line of first order
phase transitions emerges from the point
$(\infty,\kappa^{\rm f}_{\rm c},\infty)$.
But as in the Temperature-Pressure phase diagram for liquid and gas,
it ends up at a critical point in the interior of the $\beta-\kappa$ plane
(see  Fig.~\ref{phasediagram}(a)).

This analysis implies that if the semiclassical expansion around a
mean-field Ansatz yields an effective bosonic action with the matter
in the {\it fundamental} representation, there will not be any
states in the spectrum which carry the spinon quantum numbers.
Conversely,
if the matter is in the {\it adjoint} representation and
{\it breaks all the continuous symmetries}, then a phase in which there
are states that carry the fundamental color and spin-1/2 quantum numbers is
possible. Indeed, the Wilson loop in the {\it fundamental}
representation does
not need to follow the perimeter law everywhere in the interior of the
phase diagram. The dynamical matter field induced by quantum
fluctuations
carry the adjoint color charge and therefore cannot completely screen
the fundamental charge of the static sources.
Gauge invariant Green function
are {\it separately} analytic in the confinement and Higgs region.
On the upper boundary of the Higgs region, the Wilson loop
obeys the perimeter law of the deconfining regime of a pure gauge
${\cal Z}_2$ theory. By continuity, the Wilson
loop must obey the perimeter law in the Higgs region.
The same argument applies to the confinement region where now
the Wilson loop satisfies the area law. There must be a line of phase
transition separating the two regimes and connecting the classical
O(3) phase transition to the pure gauge ${\cal Z}_2$ phase transition
\cite{Fradkin 1979} (see  Fig.~\ref{phasediagram}(b)).

When we allow the norm of the matter field to fluctuate, i.e.,
$\lambda<\infty$, changes occur in the $\beta-\kappa$ phase diagram.
However, the deconfinement and Higgs regimes
always remain analytically connected for matter fields in the fundamental
representation. Conversely, the phase transition survives in
the adjoint representation \cite{Kuhnelt 1984}.

This is the important insight that the gauge picture provides us with.
The problem we are now faced with
is  how to justify the use of action
Eq.\ (\ref{fradkinaction})
where the matter is in the adjoint representations on the basis of
Eq.\ (\ref{finallatticelagrangian}). This we start now.

Let us reexamine
Eq.\ (\ref{finallatticelagrangian})
by parametrizing
the $W$'s according to
\begin{equation}
W^{\   }_{ijt}\;=\;+{\rm i}\;
\left(\;
\sqrt{|{\rm det}\; W^{\   }_{ijt}|}\;
\right)\;
V^{\   }_{ijt},\quad
V^{\   }_{ijt}\; \in\; {\rm SU(2)}.
\end{equation}
The local gauge symmetry only affects the SU(2) factor of the link degrees of
freedom. As it stands,
Eq.\ (\ref{finallatticelagrangian})
does not contain a kinetic term for the $A_{\hat{\rm o}}$'s and $V$'s.
This corresponds to the infinitely strong gauge coupling limit,
$\beta=0$. This is to be expected from gauge fields implementing a
local constraint.

So far, we have not made any approximations. However, in some
mean-field states (in particular in Wen's s-RVB) there are interesting
operators which acquire expectation values.
For instance, a term which is absent from Eq. (\ref{finallatticelagrangian}),
but which would be allowed by the symmetry, is
\begin{equation}
{\cal M}^{\   }_{ijt}\;=\;{1\over2}\;{\rm tr}\;
\left(\;
P^{\   }_{it}\;W^{\   }_{ijt}\;
P^{\dag}_{jt}\;W^{\dag}_{ijt}\;
\right),
\label{wouldbehiggskineticenergy}
\end{equation}
for any ordered pair $<i\;j>$ of lattice sites. The main point of Wen's
Ansatz is that the operator $P^{\   }_{it}$, the path ordered
product
\begin{equation}
P^{\   }_{it}\;=\;\prod_{l=1}^n\; W^{\   }_{i_{l-1}i_lt},
\quad
i_0\equiv  i,\quad i_n\equiv  i,
\end{equation}
(along the ordered pairs
$<i_0\;i_1>,\dots,<i_{l-1}\;i_l>,\dots,<i_{n-1}\;i_n>$) does acquire an
expectation value in the s-RVB ground state.
Be aware that the ${\cal M}$'s do not depend on the determinant of the $W$'s.
If we take $i$ and $j$ to be nearest-neighbor with $\epsilon$ being the
lattice constant, say $j\;=\; i\; +\; \epsilon\;\hat x$, expand $P_{jt}$ and
\begin{equation}
W^{\   }_{ijt}\;=\;
+{\rm i}\;\left(\sqrt{|{\rm det}\; W^{\   }|}\right)\;
e^{+{\rm i}\epsilon\; A^{\   }_{{\hat x}it}},\quad
A_{{\hat x}it}\;\in\;{\rm su(2)},
\end{equation}
in powers of the lattice constant, then
\begin{eqnarray}
{1\over2}
\left(
{\cal M}^{\   }_{ijt}+
{\cal M}^{*   }_{ijt}
\right)=&&
\;1
-\;{1\over4}\;{\rm tr}\;
\left(
(D_{\hat x}\; P)\;(D_{\hat x}\; P)^{\dag}
\right)
\nonumber\\
+&&\;O(\epsilon^3),
\label{adjoint}
\end{eqnarray}
where $D_{\hat x}$ is the covariant derivative in the adjoint representation,
i.e.,
\begin{equation}
D_{\hat x}\;=\;\partial_{\hat x}\;+\;{\rm i}\;[\;A_{\hat x}\;,\;\cdot\;]\;.
\end{equation}
The $P$'s can therefore be interpreted as matter fields in
the adjoint representation of SU(2) color. The fluctuations around the s-RVB
ground state should contain terms of the form of Eq. (\ref{adjoint}).
Thus, we argue that the fluctuations around an s-RVB state should behave
qualitatively like a system of matter fields in the adjoint (vector)
representation of SU(2) (which break all continuous symmetries) coupled
to SU(2) gauge fields. We are then under the general conditions of the
theorem relating the confinement and the Anderson-Higgs mechanisms
of Fradkin and Shenker \cite{Fradkin 1979}.
Notice, however, that the parameters of
the effective Lagrangian need to be determined from a microscopic
calculation. These parameters will, in turn, determine in which phase
the symmetry is realized.

In summary, we have argued how deconfinement could occur for the
fundamental
charges (the spinons) in the phase diagram of a generic lattice gauge
theory coupled to matter fields in the adjoint representation
of the gauge group.
Although Eq.\ (\ref{finallatticelagrangian}) does not resemble a canonical
SU(2) gauge-matter lattice theory, say  for example
Eq.\ (\ref{fradkinaction}),
we have identified candidates for a matter field namely the $P$'s.
Our goal in the next section
will be to construct a mean-field theory for a spin liquid
such that $P$'s are generated which break the gauge symmetry down to
${\cal Z}_2$ \cite{Wen 1991}.

\section{Mean-field theory.}
\label{sec:s-RVBMF}
\subsection{The s-RVB Ansatz.}
\label{subsec:s-RVBAnsatz}

{}From now on, we will restrict ourself to a square lattice ($d$=2) with
nearest
($J_1$) and next-nearest ($J_2$) neighbor antiferromagnetic interactions.
The s-RVB Ansatz assumes that the true ground state of the system
described by
Eq.\ (\ref{finallatticelagrangian})
is characterized by the configuration
\begin{eqnarray}
&&\bar A^{\ }_{\hat{\rm o}it}=A^{\ }_{\hat{\rm o}},
\nonumber\\
&&\bar W^{\   }_{ijt}=
\cases{
-X\sigma^3& if $j=i+\hat x$,\cr
-X\sigma^3& if $j=i+\hat y$,\cr
-{\rm Re}\ E\ \sigma^1+{\rm Im}\ E\ \sigma^2&
if $j=i+{\hat x_+}$,\cr
-{\rm Re}\ E\ \sigma^1-{\rm Im}\ E\ \sigma^2&
if $j=i+{\hat x_-}$,\cr
}
\label{sRVBAnsatzd=2}
\end{eqnarray}
of the gauge fields \cite{Wen 1991}.
The lattice spacing is taken to be one and
$\hat x\equiv (1,0)$, $\hat y\equiv (0,1)$,
$\hat x_+\equiv (+1,+1)$ and $\hat x_-\equiv (+1,-1)$.
The mean-field spectrum $\pm|\vec\xi_{\vec k}|$
for the s-RVB Ansatz is then
given by
\begin{eqnarray}
&&\xi^1_{\vec k}=
-{a^1_{\hat{\rm o}}\over2}+J_2\ {\rm Re}\ E\ \cos k_x\cos k_y,\nonumber\\
&&\xi^2_{\vec k}=
-{a^2_{\hat{\rm o}}\over2}+J_2\ {\rm Im}\ E\ \sin k_x\sin k_y,\\
&&\xi^3_{\vec k}=
-{a^3_{\hat{\rm o}}\over2}+J_1\ X\ {1\over2}\ (\cos k_x+\cos k_y).\nonumber
\end{eqnarray}

Because of the particle-hole symmetry,
the mean-field ground state is the state obtained by filling up all the
one-particle eigenstates in the Brillouin Zone $\Omega$ with negative
energy eigenvalues:
\begin{eqnarray}
|\;\Psi_{{\rm s-RVB}}>\;=\;
\prod_{\vec k,\ |\vec\xi^{\ }_{\vec k}|\leq 0}\;
\psi^{\dag}_{\vec k1}\;
\psi^{\dag}_{\vec k2}\;
|\;0\;>_{\psi}
\end{eqnarray}
where $|\ 0>_{\psi}$ is the state annihilated by all the $\psi$'s.

As shown by Wen, all the discrete symmetries of the lattice as well as parity
and time reversal are preserved by the s-RVB Ansatz when
$a^2_{\hat{\rm o}}=a^3_{\hat{\rm o}}=0$. But, most importantly to
us is that the Ansatz is designed to break all the continuous color symmetries
provided the complex mean-field parameter $E$ is neither real nor purely
imaginary and $X$ is non-vanishing.
Indeed, the counterclockwise (ccw) and clockwise (cw) products
\begin{eqnarray}
&&
P^{{\rm ccw}}_{it}\;\equiv \;
W^{\   }_{i(i+\hat x)t}\;
W^{\   }_{(i+\hat x)(i+\hat x+\hat y)t}\;
W^{\   }_{(i+\hat x+\hat y)it},\nonumber\\
&&
P^{{\rm  cw}}_{it}\;\equiv \;
W^{\   }_{i(i+\hat x)t}\;
W^{\   }_{(i+\hat x)(i+\hat x-\hat y)t}\;
W^{\   }_{(i-\hat x+\hat y)it},
\end{eqnarray}
reduce to
\begin{eqnarray}
&&
\bar P^{{\rm ccw}}\;=\;
-X^2\;\left({\rm Re}\ E\ \sigma^1-{\rm Im}\ E\ \sigma^2\right),\nonumber\\
&&
\bar P^{{\rm  cw}}\;=\;
-X^2\;\left({\rm Re}\ E\ \sigma^1+{\rm Im}\ E\ \sigma^2\right),
\end{eqnarray}
in the mean-field. Hence, $\bar P^{{\rm ccw}}$ and $\bar P^{{\rm  cw}}$
do not commute if and only if $X$, Re $E$ and Im $E$ are all non-vanishing.

The s-RVB is the phase in which $X$, Re $E$ and Im $E$ are non-vanishing.
In this phase, $|\vec\xi_{\vec k}|$ is also non-vanishing over the entire
Brillouin Zone. The spinons have acquired a gap. The spin-spin
correlation function can be calculated, within the mean-field
approximation, following the methods of Marston and Affleck
\cite{Marston 1989}. The spin-spin correlation function has an
exponential decay (in imaginary time) as a function of separation since
the s-RVB state has a gap. Also, as expected, in the BZA ($J_2/J_1 \to
0$) regime the structure factor has a peak at $(\pi,\pi)$ and, in the
flux regime ($J_2/J_1 \geq (J_2/J_1)_{\rm cr}$)
it has peaks at $(\pi/2,\pi/2)$,
$(0,\pi)$ and $(\pi,0)$.

\subsection{The saddle-point equations and mean-field ground states.}
\label{subsec:MFgroundstates}

The saddle-point equations in the infinite volume limit are
\begin{equation}
0=F_i(\{\kappa\}),\qquad i=1,\cdots,6,
\end{equation}
where $\{\kappa\}$ is the set made of the mean-field parameters
\begin{eqnarray}
&&
\kappa_1= a_{\hat{\rm o}}^1,\quad
\kappa_2= a_{\hat{\rm o}}^2,\quad
\kappa_3= a_{\hat{\rm o}}^3,\nonumber\\
&&
\kappa_4= {\rm Re}\ E,\quad
\kappa_5= {\rm Im}\ E,\quad
\kappa_6=         \ X,
\end{eqnarray}
and the functions $F_i$ are defined implicitly by integrals over a
square with the area $4\pi^2$ (the Brillouin Zone $\Omega$):
\begin{eqnarray}
F_1&&={1\over4\pi^2}\int_{\Omega}\ {\xi_1\over|\vec\xi|},\quad
F_4  ={1\over4\pi^2}\int_{\Omega}\ {\xi_1\over|\vec\xi|}\ g_1\ -\ \kappa_4,
\nonumber\\
F_2&&={1\over4\pi^2}\int_{\Omega}\ {\xi_2\over|\vec\xi|},\quad
F_5  ={1\over4\pi^2}\int_{\Omega}\ {\xi_2\over|\vec\xi|}\ g_2\ -\ \kappa_5,
\nonumber\\
F_3&&={1\over4\pi^2}\int_{\Omega}\ {\xi_3\over|\vec\xi|},\quad
F_6  ={1\over4\pi^2}\int_{\Omega}\ {\xi_3\over|\vec\xi|}\ g_3\ -\ \kappa_6.
\end{eqnarray}
Here,
\begin{eqnarray}
&&g_1(\vec k)\;=\;\cos k_x\; \cos k_y,\nonumber\\
&&g_2(\vec k)\;=\;\sin k_x\; \sin k_y,\nonumber\\
&&g_3(\vec k)\;=\;{1\over2}\;(\cos k_x\ +\ \cos k_y),
\end{eqnarray}
each transform irreducibly under the point group of the square lattice.
Hence, the integrals only need to be performed over the reduced Brillouin
zone $\Omega'\ =\ \{\vec k\ \in\ \Omega\ |\ 0\leq|k_y|\leq\pi\}$.
It is also useful to consider the Jacobian
\begin{equation}
F_{i,j}=
{\partial F_i\over \partial \kappa_j}\qquad i,j=1,\cdots,6.
\end{equation}

There are two limiting cases for which the saddle-point equations can be
solved analytically. The BZA limit is the limit when $J_2=0$
\cite{Baskaran 1987}.
The mean-field parameters
\begin{equation}
\kappa_1=
\kappa_2=
\kappa_3=
\kappa_4=
\kappa_5=0,\quad
\kappa_6={4\over\pi^2},
\end{equation}
are (up to a sign) the unique solutions of the saddle-point equations.
The single particle excitation spectrum is that of a tight-binding
fermionic gas at half-filling. The edges of the square with vertices
$(\pi,0)$, $(0,\pi)$, $(-\pi,0)$, $(0,-\pi)$ at which the spectrum
is gapless is the Fermi surface. The Jacobian is formally given by the matrix
\begin{equation}
\pmatrix
{
F_{1,1}&0&0&0&0&0\cr
0&F_{2,2}&0&0&0&0\cr
0&0&0&      0&0&0\cr
0&0&0&F_{4,4}&0&0\cr
0&0&0&0&F_{5,5}&0\cr
0&0&0&0&0&F_{6,6}\cr
}_{{\rm BZA}}.
\end{equation}
It turns out that the non-vanishing entries are divergent integrals.

The limiting case $J_1=0$ can also be solved analytically.
The mean-field parameters
\begin{eqnarray}
&&
\kappa_1=
\kappa_2=
\kappa_3=
\kappa_6=0,\nonumber\\
&&
\kappa_4=\kappa_5=
{1\over4\pi^2}{1\over2}\int_{\Omega}\sqrt{g_1^2+g_2^2}\approx
0.339,
\end{eqnarray}
are (up to signs) the unique solutions of the saddle-point equations.
The single particle excitation spectrum has four discrete gapless points
on the Brillouin Zone. They are located at
\begin{eqnarray}
&&
\vec k_{01}\;=\;(\;+{\pi\over2}\;,\;0\;),\quad
\vec k_{02}\;=\;(\;0\;,\;+{\pi\over2}\;),\quad
\nonumber\\
&&
\vec k_{03}\;=\;(\;-{\pi\over2}\;,\;0\;),\quad
\vec k_{04}\;=\;(\;0\;,\;-{\pi\over2}\;).
\label{fluxminima}
\end{eqnarray}
In the neighborhood of each of these nodes, the
excitation spectrum takes the same form as that of a relativistic massless
particle. It is shown in appendix \ref{sec:flux} that
this gapless state is characterized
by an average mean-field flux of $\pi$ through a cell bounded by the
next-nearest-neighbor links and hence it
will be called the flux state \cite{AMfluxstate}.
In this limit, the Jacobian is well defined and given by
\begin{equation}
\pmatrix
{
F_{1,1}&0&0&0&0&0\cr
0&F_{2,2}&0&0&0&0\cr
0&0&F_{3,3}&0&0&0\cr
0&0&0&F_{4,4}&F_{4,5}&0\cr
0&0&0&F_{5,4}&F_{5,5}&0\cr
0&0&0&0&0&F_{6,6}\cr
}_{{\rm flux}}.
\end{equation}

For arbitrary values of the ratio $J_2/J_1$, the saddle-point equations
can only be solved numerically. The value of the solutions of the
saddle-point equations as a function of $J_2/J_1$ are plotted
in  Fig.~\ref{mfsolutions}.
One observes that $\kappa_2=\kappa_3=0$. This results from
the lattice symmetry.
It also appears that there exists a critical value
for the ratio $J_2/J_1$ above which the flux mean-field solution
is always preferred. The numerics predict this critical value to be
between 1.3 and 1.4.

This critical value can be found analytically by requiring the Jacobian
determinant to satisfy
\begin{equation}
{\rm det}\
\left({\partial F_i\over\partial\kappa_j}\right)_{{\rm flux}}=0.
\end{equation}
It is understood here that the Jacobian is calculated for the mean-field
parameters in the flux limit but for arbitrary values of the ratio
$J_2/J_1$. This condition is thus simply
\begin{equation}
{J_2\over J_1}=
{1\over4\pi^2}{1\over\kappa_4}\int_{\Omega} {g_3^2\over\sqrt{g_1^2+g_2^2}}
\approx 1.342
\end{equation}
if $\kappa_4\approx 0.339$.

The numerics close to the BZA regime are too unreliable to predict the
existence of a critical value of the ratio $J_2/J_1$ below which the
BZA state prevails. The analytical argument
used for the flux phase fails as well in this limit
since the Jacobian is not of maximal rank and moreover is ill defined.
Indeed, singularities associated with the Fermi surface renders all
non-vanishing entries of the Jacobian infinite.

The numerical results clearly indicate that, between $J_2/J_1=0.6$
and $J_2/J_1=0.7$, the topography of the single particle energy crosses over
from a regime with the absolute minima located on the diagonal of $\Omega'$
to a regime with the absolute minima located on the edges of
the reduced Brillouin Zone (see  Fig.~\ref{spinonenergysurface}).
There should therefore exists a crossover value for the ratio $J_2/J_1$
for which there is an accidental degeneracy of the minima.
The Lagrange multiplier $|a^1_{\hat{\rm o}}|$
appears to keep track of this crossover, the point at which it peaks.
Our mean-field results are summarized in  Fig.~\ref{MFRG}.

\subsection{Higher dimensional generalization.}
\label{subsec:MFd>2}

We will try in this section to generalize the s-RVB Ansatz to any
spatial dimension $d>2$. Let $\Lambda={\cal Z}^d$
be a $d$ dimensional hypercubic
lattice on which Eq. (\ref{finallatticelagrangian}) is defined.
It is helpful to introduce the $2d$ unit vectors
\begin{eqnarray}
&&
\hat x^{\mu}_{\sigma_{\mu}}\equiv
(0\;,\cdots,\;0\;,\;\sigma_{\mu}\;,\;0\;,\cdots,\;0),
\nonumber\\
&&
\mu=1,\cdots,d,\quad\sigma_{\mu}=\pm1
\end{eqnarray}
and the $2d(d-1)$ vectors of norm $\sqrt{2}$
\begin{eqnarray}
&&
\hat x^{\mu\nu}_{\sigma_{\mu\nu}}\equiv
(0\;,\cdots,\;0\;,\;\sigma_{\mu}\;,\;0\;,\cdots,
\;0\;,\;\sigma_{\nu}\;,\;0\;,\cdots,\;0),
\nonumber\\
&&
\mu,\nu=1,\cdots,d,\quad
\sigma_{\mu\nu}\equiv \sigma_{\mu}\cdot\sigma_{\nu},\quad
\mu<\nu,
\end{eqnarray}
which connect any given site with all its nearest-neighbors
and next-nearest-neighbors, respectively.
The straightforward generalization of Eq. (\ref{sRVBAnsatzd=2})
is
\begin{eqnarray}
&&\bar A^{\ }_{\hat{\rm o}it}=A^{\ }_{\hat{\rm o}},
\nonumber\\
&&\bar W^{\   }_{ijt}=
\cases{
-X\sigma^3& if $j=i+\hat x^{\mu}_{\sigma_{\mu}}$,\cr
-{\rm Re}\ E\ \sigma^1+{\rm Im}\ E\ \sigma^2&
if $j=i+\hat x^{\mu\nu}_{+1}$,\cr
-{\rm Re}\ E\ \sigma^1-{\rm Im}\ E\ \sigma^2&
if $j=i+\hat x^{\mu\nu}_{-1}$.\cr
}
\label{sRVBAnsatzd>2}
\end{eqnarray}

The breaking of all the continuous color symmetries is automatically
achieved once this is so for $d=2$. However,
we need to check that the Ansatz is, up to gauge transformations,
invariant under all point group transformations of $\Lambda$
as well as under time reversal.

Time reversal is equivalent to complex conjugation.
In turn, complex conjugation is equivalent to
${\rm Im}\ E\rightarrow-{\rm Im} E$. But, this change can be undone by
the gauge transformation
\begin{equation}
U_{i}\;\equiv \;{\rm sgn}(i)\;\sigma^1,
\label{generatortimereversal}
\end{equation}
provided $A_{\hat{\rm o}}\;=\;{1\over2}\;a^1_{\hat{\rm o}}\;\sigma^1$.
Here, sgn($i$) is the lattice function defined in Eq. (\ref{latticesign}).

We now turn to the point group transformations of $\Lambda$.
They are generated cyclically from the $d!$ permutations of the
coordinates of $i\in\Lambda$:
\begin{eqnarray}
(i_1\;,\;\cdots\;,\;i_d)\rightarrow
(i_{n_1}\;,\;\cdots\;&&,\;i_{n_d}),
\nonumber\\
&&
n_1,\cdots,n_d=1,\cdots,d,
\label{d!permutations}
\end{eqnarray}
and from the $d$ transformations
\begin{eqnarray}
&&
(i_1\;,\;\cdots\;,\;i_d)\rightarrow
(-i_1\;,\;\cdots\;,\;i_d),\cdots
\nonumber\\
&&\qquad\qquad
\cdots,
(i_1\;,\;\cdots\;,\;i_d)\rightarrow
(i_1\;,\;\cdots\;,\;-i_d).
\label{dsignchanges}
\end{eqnarray}

When $d=2$, the $d!=2$ permutations are the identity and the
reflection about the diagonal $x=y$. They leave
Eq. (\ref{sRVBAnsatzd>2}) unchanged. The remaining point group
transformations are generated by reflections about the $x$ or $y$ axis.
For $d=2$, they are equivalent to complex conjugation and Wen's assertion
that Eq. (\ref{sRVBAnsatzd>2}) describes a spin liquid follows.
But for higher dimensions, the equivalence between complex conjugation
and any of the $d$ reflections about a $d-1$ dimensional hyperplane,
Eq. (\ref{dsignchanges}), is not true anymore.

A possible cure to this problem is to find a partitioning of all the
next-nearest-neighbor links into two disjoint families which are
invariant under all point group transformations. Notice first that
the lattice function
\begin{equation}
{\rm sgn}:\;\Lambda\rightarrow{\cal Z}_2,\;
i\rightarrow{\rm sgn}(i)=(-1)^{i_1+\cdots +i_d}
\label{latticesign}
\end{equation}
is unchanged by any permutation of the coordinates of $i$.
Furthermore, sgn($i$) is also left unchanged when any one
of the coordinates of $i$ has its sign changed. Consequently,
the partitioning of $\Lambda$ into the even (odd)
sites $\Lambda^{\ }_{\rm e (o)}$ is not affected by the point
group transformations.

We immediately see that the Ansatz
\begin{eqnarray}
&&\bar A^{\ }_{\hat{\rm o}it}=A^{\ }_{\hat{\rm o}},
\label{sRVBAnsatzdgeq2}
\\
&&\bar W^{\   }_{ijt}=
\cases{
-X\sigma^3& if $j=i+\hat x^{\mu}_{\sigma_{\mu}}$,\cr
-{\rm Re}\ E\sigma^1+{\rm Im}\ E\ \sigma^2&
if $j=i+\hat x^{\mu\nu}_{\pm1},\;i\in\Lambda_{\rm e}$,\cr
-{\rm Re}\ E\sigma^1-{\rm Im}\ E\ \sigma^2&
if $j=i+\hat x^{\mu\nu}_{\pm1},\;i\in\Lambda_{\rm o}$,\cr
}
\nonumber
\end{eqnarray}
respects all the point group symmetries.
On the other hand, translational invariance and time reversal
appear to be broken.
However, they affect the Ansatz
Eq. (\ref{sRVBAnsatzdgeq2})
in the same way as the gauge transformation
Eq. (\ref{generatortimereversal}) does.
We conclude that this Ansatz describes a spin liquid which breaks all
the continuous symmetries of color SU(2).

In two spatial dimensions, this new Ansatz reduces to the BZA Ansatz
in {\it both} the $J_2/J_1\rightarrow0$ and $J_1/J_2\rightarrow0$ limits.
Moreover, it is not possible to construct a mean-field plaquette term
threaded by a flux of $\pi$.
We therefore expect that a gap to the mean-field spinon excitations is
generated for all $0<J_1/J_2<\infty$.

\section{Gauge fields fluctuations close to and in the flux phase.}
\label{sec:fluc}

It is fortunate that the mean-field theory yields the flux
phase since one can relatively easily identify a continuum effective
theory for the long wavelength fluctuations of the soft modes.
In contrast, it is much harder to do the same close to the BZA limit.
This is so because in the flux phase the locus of points
in the Brillouin Zone for which the mean-field single particle energy
reaches a minima are isolated. From the location of the isolated minima,
one can then extract a reciprocal vector around which the Fourier transform of
the spinon and link degrees of freedom can be expanded, thus yielding
the intermediate continuum limit.
However, it is not feasible to carry out such an approach
at the BZA point since there there are infinitely many degenerate minima
along a square, {\it the spinon Fermi surface}, with vertices
$(\pm\pi,0)$ and $(0,\pm\pi)$ in the Brillouin Zone.
Moreover, the difficulty is compounded by the existence of saddle points
in the energy surface at the vertices of the spinon Fermi surface
causing Van-Hove singularities in the mean-field density of states
and by the nesting property of the Fermi surface.
One possible drastic approximation is to replace the anisotropic Fermi surface
altogether by a circle and then expand the nearest-neighbor
links degrees of freedom in powers of the gauge fields up to second order.
The continuum effective theory is that of gauge fields coupled to
non-relativistic fermions. But, by doing so, one neglects magnetic
instabilities induced by the nesting property of the Fermi surface
at half-filling. Away from half-filling, such an approximation has been
used by Lee and Nagaosa \cite{Lee 1992} in their study of the $t-J$ model.
It is then known that the fluctuations of the gauge fields, when coupled to
the ``Fermi surface'', lead to severe infrared divergences
\cite{FLinstability} which possibly signal an instability of the
mean-field theory (and thus of the ``Fermi surface'') to a
``strange metal'' \cite{Lee 1992}.
Given these difficulties, we will concentrate on the flux phase and
will not discuss the BZA regime any further. In any event, the
mean-field theory predicts that the BZA and flux regimes are smoothly
connected. Thus, we expect that a detailed study of the flux regime can
throw light on the behavior near the BZA limit as well. Notice, however,
that the low energy fermionic states have different physical
properties in both regimes given
the level crossing of the excited states shown in the previous section.
Nevertheless, since the mean-field ground state evolves smoothly (i.e.,
without level crossings) we will still assume that the physics
is qualitatively similar at both ends.

In the spirit of section \ref{sec:Deconfinement},
we need to identify the bosonic degrees of freedom of the continuum limit.
To this end we will perform a semiclassical expansion of the link degrees
of freedom around the mean-field Ansatz.
In the language of critical phenomena,
we will use the fact that the mean-field theory of the previous
section  predicts a second order phase transition at
$(J_2/J_1)_{{\rm cr}}$. We will then think of the
continuum limit as the tuning of the coupling $J_2/J_1$ toward the point
$(J_2/J_1)_{{\rm cr}}$ where a correlation length
(the inverse of the spinon mass) becomes infinite.
In this scheme, the intermediate effective field theory we
obtain is the (infrared unstable) fixed point action of
Eq. (\ref{finallatticelagrangian}) in the
sense of the renormalization group
(see  Fig.~\ref{MFRG}). Notice that the semiclassical
expansion is not fully justified since the theory, as it stands, does
not have a small parameter that will control the fluctuations. Formally,
we can regard our semiclassical expansion as a $1/N$ expansion in
which $N$ has been set to unity. While this approach seems tempting, we
should be aware that at small values of $N$ the physics may be different
from the semiclassical (mean-field) picture.

The intermediate effective theory so obtained will contain gauge fields
and matter fields, which both transform like the adjoint (vector)
representation  of SU(2) color, and are coupled with (several species of)
Dirac fermions.
However, the bosons are still not truly dynamical since they do not possess a
kinetic energy of their own. On the other hand, all the symmetry requirements
that we are after are fulfilled, namely the bosonic fields break all
the continuous gauge symmetries down to a discrete subgroup ${\cal
Z}_2$. In the next subsections we derive the effective continuum theory
in the vicinity of the flux phase. It turns out that Wen's choice of
gauge is not very natural in the flux regime. In appendix \ref{sec:flux}
we give a
more convenient construction in a different (``flux") gauge which makes
simpler the description of the fluctuations around this state. The
reader is referred to appendix \ref{sec:flux}
for the notation that we will use below.

\subsection{Continuum limit in the flux phase when $J_1=0$.}
\label{subsec:J1=0}

We will assume that $J_1$ has been turned off. The fermions on the even
sublattice are then decoupled from those on the odd sublattice. We shall
show how to take the continuum limit so as to obtain relativistic
Dirac spinors in 2+1 dimensions which couple via the minimal coupling to
two independent set of SU(2) gauge fields.

We will choose to parametrize small
fluctuations of the $Q$'s (see Eq. (\ref{allnextnearestneighborlinks}))
around the mean-field Ansatz by
\begin{eqnarray}
Q^{e1}_{\hat x_-i}=&&+{\rm i}|E|\
e^{-{\rm i}\epsilon'A^{e1}_{\hat x_-i}},\quad
Q^{e1}_{\hat x_+i}=  -{\rm i}|E|\
e^{-{\rm i}\epsilon'A^{e1}_{\hat x_+i}},
\nonumber\\
Q^{e2}_{\hat x_-i}=&&+{\rm i}|E|\
e^{-{\rm i}\epsilon'A^{e2}_{\hat x_-i}},\quad
Q^{e2}_{\hat x_+i}=  +{\rm i}|E|\
e^{-{\rm i}\epsilon'A^{e2}_{\hat x_+i}},
\nonumber\\
Q^{e3}_{\hat x_-i}=&&+{\rm i}|E|\
e^{-{\rm i}\epsilon'A^{e3}_{\hat x_-i}},\quad
Q^{e3}_{\hat x_+i}=  +{\rm i}|E|\
e^{-{\rm i}\epsilon'A^{e3}_{\hat x_+i}},
\nonumber\\
Q^{e4}_{\hat x_-i}=&&+{\rm i}|E|\
e^{-{\rm i}\epsilon'A^{e4}_{\hat x_-i}},\quad
Q^{e4}_{\hat x_+i}=  -{\rm i}|E|\
e^{-{\rm i}\epsilon'A^{e4}_{\hat x_+i}}.
\nonumber\\
&&
\label{nonlinearparametrization}
\end{eqnarray}
Here, $\epsilon'(\bar\epsilon)$ is related to the lattice constant
$\epsilon$ by
\begin{equation}
\epsilon'\;=\;\sqrt{2}\;\epsilon,\quad (\;\bar\epsilon\;=\;2\;\epsilon'\;).
\end{equation}
Only SU(2) color fluctuations of the link degrees of freedom are considered
since the determinant $|E|$ is kept unchanged.

In the Heisenberg
representation, the equations of motion satisfied by the fermions are,
to lowest order in $\bar\epsilon$,
\begin{eqnarray}
{\rm i}D^{e1}_{\hat{\rm o}}f^{e1}_{i}=&&
{\rm v}_{\rm f}
\Bigg[
\left(
{\rm i}\partial_{\hat x_-}
+{1\over2}(A^{e1}_{\hat x_-i}+A^{e2}_{\hat x_-i})
\right)
f^{e2}_{i}
\nonumber\\
-&&
\left(
{\rm i}\partial_{\hat x_+}
+{1\over2}(A^{e1}_{\hat x_+i}+A^{e4}_{\hat x_+i})
\right)
f^{e4}_{i}
\Bigg],
\nonumber\\
{\rm i}D^{e2}_{\hat{\rm o}}f^{e2}_{i}=&&
{\rm v}_{\rm f}
\Bigg[
\left(
{\rm i}\partial_{\hat x_-}
+{1\over2}(A^{e2}_{\hat x_-i}+A^{e1}_{\hat x_-i})
\right)
f^{e1}_{i}
\nonumber\\
+&&
\left(
{\rm i}\partial_{\hat x_+}
+{1\over2}(A^{e2}_{\hat x_+i}+A^{e3}_{\hat x_+i})
\right)
f^{e3}_{i}
\Bigg],
\nonumber\\
{\rm i}D^{e3}_{\hat{\rm o}}f^{e3}_{i}=&&
{\rm v}_{\rm f}
\Bigg[
\left(
{\rm i}\partial_{\hat x_-}
+{1\over2}(A^{e3}_{\hat x_-i}+A^{e4}_{\hat x_-i})
\right)
f^{e4}_{i}
\nonumber\\
+&&
\left(
{\rm i}\partial_{\hat x_+}
+{1\over2}(A^{e3}_{\hat x_+i}+A^{e2}_{\hat x_+i})
\right)
f^{e2}_{i}
\Bigg],
\nonumber\\
{\rm i}D^{e4}_{\hat{\rm o}}f^{e4}_{i}=&&
{\rm v}_{\rm f}
\Bigg[
\left(
{\rm i}\partial_{\hat x_-}
+{1\over2}(A^{e4}_{\hat x_-i}+A^{e3}_{\hat x_-i})
\right)
f^{e3}_{i}
\nonumber\\
-&&
\left(
{\rm i}\partial_{\hat x_+}
+{1\over2}(A^{e4}_{\hat x_+i}+A^{e1}_{\hat x_+i})
\right)
f^{e1}_{i}
\Bigg].
\end{eqnarray}
Here, the flux velocity ${\rm v}_{\rm f}$ is
\begin{equation}
{\rm v}_{\rm f}\;=\;{J_2\;|E|\;\bar\epsilon\over4},
\end{equation}
and the covariant time derivative is
\begin{equation}
D^{ea}_{\hat{\rm o}}=\partial_t+{\rm i}A^{ea}_{\hat{\rm o}i},\quad
a=1,2,3,4.
\end{equation}
The same equations hold on $\Lambda^{(1)}_{\rm o}$.

In the continuum limit, we make the following identifications:
\begin{eqnarray}
&&
\lim_{\bar\epsilon\rightarrow0}\
{f^{e1}_i+f^{e2}_i\over\bar\epsilon}\equiv
u^1(\vec rt),
\quad
\lim_{\bar\epsilon\rightarrow0}\
{f^{e3}_i+f^{e4}_i\over\bar\epsilon}\equiv
v^1(\vec rt),
\nonumber\\&&
\lim_{\bar\epsilon\rightarrow0}\
{f^{e3}_i-f^{e4}_i\over\bar\epsilon}\equiv
u^2(\vec rt),
\quad
\lim_{\bar\epsilon\rightarrow0}\
{f^{e2}_i-f^{e1}_i\over\bar\epsilon}\equiv
v^2(\vec rt),
\nonumber\\&&
\lim_{\bar\epsilon\rightarrow0}\
A^{ea}_{\hat{\rm x}_-i}\equiv  A^e_1(\vec rt),
\quad
\lim_{\bar\epsilon\rightarrow0}\
A^{ea}_{\hat{\rm x}_+i}\equiv  A^e_2(\vec rt),
\nonumber\\
&&
\lim_{\bar\epsilon\rightarrow0}\
{A^{ea}_{\hat{\rm o}i}\over{\rm v}_{\rm f}}
\equiv  A^e_0(\vec rt),
\quad a=1,2,3,4,
\nonumber\\
&&
\lim_{\bar\epsilon\rightarrow0}\
\bar\epsilon^2\sum_{i\in\Lambda^{(1)}_{\rm e}}\equiv
\int d^2r,
\end{eqnarray}
on $\Lambda^{(1)}_{\rm e}$.
The counterparts to the fields $u$, $v$ and $A^e_{\mu}$
on sublattice $\Lambda^{(1)}_{\rm o}$ are $w$, $z$ and $A^o_{\mu}$,
respectively.

The equations of motion, when rewritten in terms of
\begin{equation}
u=\pmatrix{u^1\cr u^2},\quad
v=\pmatrix{v^1\cr v^2},\quad
w=\pmatrix{w^1\cr w^2},\quad
z=\pmatrix{z^1\cr z^2},
\end{equation}
take the same form as those for four independent massless
Dirac spinors in 2+1 space-time dimensions, the speed of light
being here the flux velocity. To describe the relativistic
structure, we need the three gamma matrix
\begin{eqnarray}
&&
\gamma^\mu\equiv
(\tau^2\tau^0,-\tau^2\tau^3,-\tau^2\tau^1)=
(\tau^2,-{\rm i}\tau^1,+{\rm i}\tau^3),
\nonumber\\
&&
\not\!\!D^{e(o)}_{\mu}\equiv
\gamma^{\mu}D^{e(o)}_{\mu}\equiv
\gamma^{\mu}
\left(
\partial_{\mu}+{\rm i}A^{e(o)}_{\mu}
\right),
\end{eqnarray}
where the $\vec\tau$ are a new set of Pauli matrices.
The long wavelength, low energy limit of
Eq.\ (\ref{finallatticelagrangian})
when $J_1=0$ and for gauge fields close to the flux phase configuration
is
\begin{eqnarray}
&&
L=\int d^2r\; {\cal L},
\label{continummlimitJ1=0}\\
&&
{\cal L}=
{\rm v}_{\rm f}
\left[
\bar u\ {\rm i}\not\!\! D^e\ u+
\bar w\ {\rm i}\not\!\! D^o\ w+
\bar v\ {\rm i}\not\!\! D^e\ v+
\bar z\ {\rm i}\not\!\! D^o\ z
\right].
\nonumber
\end{eqnarray}
{\it
The dynamics due to fluctuations in the determinant of the $W$'s has been
neglected
} \cite{determinantfluctuations}.

There are thus four flavors of Dirac spinors which interact with
two independent set of SU(2) color gauge fields through the minimal coupling.
In the continuum limit, the original $16=2\times(4+4)$ fermionic
degrees of freedom per unit cell of
$\Lambda^{(1)}_{\rm e}\cup\Lambda^{(1)}_{\rm o}$
have become $16=4\times2\times2$
fields carrying flavor, color and spinor indices, respectively.
On the other hand, the original $16=2\times4+2\times4$
degrees of freedom residing on the oriented links of the unit cell
have been projected onto the smaller manifold
of $4=2+2$ su(2) spatial gauge fields.
What we have done could have been achieved by expanding the equations of
motion in reciprocal space around the minima of the mean-field
spinon spectrum and keeping only the lowest terms in powers of the
reciprocal lattice. The gauge fields that are neglected vary rapidly
compared to the one kept. Choosing a staggered gauge
transformation is equivalent to choosing the origin in reciprocal space to
be the location of one of the degenerate minima. The different flavors
of the continuum spinons is caused by the degeneracy of the mean-field
spectrum.

The local gauge symmetry of
Eq.\ (\ref{continummlimitJ1=0})
is that of the semi-simple gauge group SU(2)$\times$SU(2).
Indeed, if we arrange the fermions and gauge fields
into the $(j_1,j_2)=(1/2,1/2)$ reducible representation of
SU(2)$\times$SU(2) \cite{representation}
\begin{equation}
\Psi\equiv
\pmatrix
{
u\cr
w\cr
},
\quad
\Theta\equiv
\pmatrix
{
v\cr
z\cr
},
\quad
A_{\mu}\equiv
\pmatrix
{
A^e_{\mu}&0\cr
0&A^o_{\mu}\cr
},
\end{equation}
then
\begin{equation}
{\cal L}=
{\rm v_f}
\left[
\bar\Psi   \ {\rm i}\not\!\! D\ \bar\Psi  +
\bar\Theta \ {\rm i}\not\!\! D\ \bar\Theta
\right]
\label{L0}
\end{equation}
is invariant under the local gauge transformation
\begin{eqnarray}
&&
\Psi(\Theta)    \rightarrow U\ \Psi(\Theta),\quad
A_{\mu}\ \rightarrow U\ A_{\mu}\ U^{\dag}\
+\ ({\rm i}\partial_{\mu}\ U)\ U^{\dag},
\nonumber\\
&&
\forall U\equiv
\pmatrix
{
U^e&0\cr
0&U^o\cr
}\
\in\
{\rm SU(2)}\times{\rm SU(2)}.
\end{eqnarray}

\subsection{\bf Continuum limit in the flux phase when $J_1\neq0$.}
\label{subsec:J1neq0}

The fermions on the even and odd sublattices are coupled by pure
fluctuations of the degrees of freedom $M$,
Eq.\ (\ref{allnearestneighborlinks}),
on the nearest-neighbor links.
These fluctuations are of order $O(\epsilon)$ and therefore it is
meaningless to separate the contributions from the SU(2) color and
determinant factors \cite{dimerinstability}.
Thus, the fluctuating $M$ form a four dimensional real vector space.
By repeating the same steps as in the previous subsection,
one finds the long wavelength, low energy continuum limit of
Eq.\ (\ref{finallatticelagrangian})
to be given by
\begin{equation}
{\cal L}\;=\;{\cal L}_0\;+\;{\cal L}_1\;+\;{\cal L}_2.
\label{Linfluxphase}
\end{equation}
The first contribution ${\cal L}_0$ is the one given by
Eq.\ (\ref{L0}) in the limit $J_1=0$. The second and third contribution
account for the fluctuations of the $M$ on the nearest-neighbor links.
There are 16 such degrees of freedom.

Four of them, $N^{eo}_1,\dots, N^{eo}_4$, transform like scalars
under Lorentz transformations. They are linearly independent
combinations of the original lattice degrees of freedom connecting even to odd
sites and have been rescaled so has to carry the dimensions of inverse length.
With the help of the four dimensional representation
\begin{eqnarray}
{\cal N}^{\   }_1\equiv
\pmatrix
{
0& {\rm i}N^{eo\   }_1\cr
-{\rm i}N^{eo\dag}_1&0\cr
},\
{\cal N}^{\   }_2\equiv
\pmatrix
{
0& {\rm i}N^{eo\   }_2\cr
-{\rm i}N^{eo\dag}_3&0\cr
},
\nonumber\\
{\cal N}^{\   }_3\equiv
\pmatrix
{
0& {\rm i}N^{eo\   }_3\cr
-{\rm i}N^{eo\dag}_2&0\cr
},\
{\cal N}^{\   }_4\equiv
\pmatrix
{
0& {\rm i}N^{eo\   }_4\cr
-{\rm i}N^{eo\dag}_4&0\cr
},
\nonumber\\
&&\hskip -1true cm
\label{thecalN}
\end{eqnarray}
the second contribution to
Eq.\ (\ref{Linfluxphase})
is
\begin{eqnarray}
&&
{\cal L}_1=
-{J_1\over16}\ \sum_{n=1}^4\
{\rm tr}\
\left[
{\cal N}^{\   }_n\ {\cal N}^{\dag}_{n}
\right]
\\
&&\ +
{J_1\ {\rm v_f}\over\sqrt{2}J_2|E|}
\left[
\bar\Psi   \ {\cal N}^{\   }_1\ \Psi   +
\bar\Psi   \ {\cal N}^{\   }_2\ \Theta +
\bar\Theta \ {\cal N}^{\   }_3\ \Psi   +
\bar\Theta \ {\cal N}^{\   }_4\ \Theta
\right].
\nonumber
\end{eqnarray}

Notice the mixing of the four Dirac species (or flavors).
Under a local SU(2)$\times$SU(2) color gauge transformation $U$,
the ${\cal N}$'s transform like the adjoint representation:
\begin{equation}
{\cal N}^{\   }_n\;\rightarrow\; U^{\   }\; {\cal N}^{\   }_n\; U^{\dag},
\quad n=1,\dots,4.
\end{equation}
As it turns out, the ${\cal N}$'s pick up mean-field expectation
values below the
flux phase which break the SU(2)$\times$SU(2) symmetry down to ${\cal Z}_2$,
the center of SU(2) color.

The description of the remaining twelve degrees of freedom follows the same
line as that for the ${\cal N}$'s except that they can be grouped into
four families,
$A^{eo}_{\mu}$,
$B^{eo}_{\mu}$,
$C^{eo}_{\mu}$,
$D^{eo}_{\mu}$,
each transforming like a three-vector under Lorentz transformation.
With the help of the four dimensional representation
\begin{eqnarray}
{\cal A}^{\   }_{\mu}\equiv
\pmatrix
{
0&{A}^{eo\   }_{\mu}\cr
{A}^{eo\dag}_{\mu}&0\cr
},\quad
{\cal B}^{\   }_{\mu}\equiv
\pmatrix
{
0&{B}^{eo\   }_{\mu}\cr
{C}^{eo\dag}_{\mu}&0\cr
},
\nonumber\\
{\cal C}^{\   }_{\mu}\equiv
\pmatrix
{
0&{C}^{eo\   }_{\mu}\cr
{B}^{eo\dag}_{\mu}&0\cr
},\quad
{\cal D}^{\   }_{\mu}\equiv
\pmatrix
{
0&{D}^{eo\   }_{\mu}\cr
{D}^{eo\dag}_{\mu}&0\cr
},
\nonumber\\
\label{thecalABCD}
\end{eqnarray}
the third contribution to
Eq.\ (\ref{Linfluxphase})
is
\begin{eqnarray}
&&
{\cal L}_2=
-{J_1\over16}\
{\rm tr}\
\left[
{\cal A}^{\mu }\ {\cal A}^{\dag}_{\mu}+
{\cal B}^{\mu }\ {\cal B}^{\dag}_{\mu}+
{\cal C}^{\mu }\ {\cal C}^{\dag}_{\mu}+
{\cal D}^{\mu }\ {\cal D}^{\dag}_{\mu}
\right]
\nonumber\\
&&\ +
{J_1\ {\rm v_f}\over\sqrt{2}J_2|E|}
\left[
\bar\Psi\    \not\!\! {\cal A}^{\   }\ \Psi   +
\bar\Psi\    \not\!\! {\cal B}^{\   }\ \Theta +
\bar\Theta \ \not\!\! {\cal C}^{\   }\ \Psi   +
\bar\Theta \ \not\!\! {\cal D}^{\   }\ \Theta
\right].
\nonumber\\
\end{eqnarray}
The structure of ${\cal L}_2$ is the same as that for ${\cal L}_1$
except for the relativistic transformation properties of the dynamical fields.
Any mean-field expectation value of the ${\cal A}_{\mu},\dots,{\cal D}_{\mu}$'s
below the flux phase would imply a loss of relativistic covariance.
But, as it turns out, these fields do not pick up a mean-field
expectation value just below the onset of the flux phase.
We do not expect that they play a relevant role to the issue of the
confinement of the spinons.

\subsection{Continuum limit just below $(J_2/J_1)_{\rm cr}$.}
\label{subsec:J2<J2cr}

Below the onset of the flux phase, the mean-field solutions of
the saddle-point equations take the form
\begin{eqnarray}
&&
a^{1}_{\hat{\rm o}}\neq 0,\quad
a^{2}_{\hat{\rm o}}=    0,\quad
a^{3}_{\hat{\rm o}}=    0,
\\
&&
{\rm Re}\ E\neq 0,\quad
{\rm Im}\ E\neq 0,\quad
X\neq 0.
\nonumber
\end{eqnarray}
The six mean-field parameters evolve continuously into their
flux phase values as $J_2/J_1$ approaches from below its critical value.
What continuum limit can we expect close to the onset of the flux phase?
Strictly speaking, the locations of the minima for
the mean-field excitation spectrum are functions of the parameters
$\vec a_{\hat{\rm o}}$, $E$ and $X$, i.e., of $J_2/J_1$ and it would appear
that any analytical approach is hopeless.

However, let us expand formally the mean-field energy excitations
in powers of the lattice spacing $\epsilon$ by inserting
\begin{equation}
\vec k_l=\vec k_{0l}+\epsilon\vec p,
\quad l=1,2,3,4,
\end{equation}
into the mean-field dispersion relation. Here, the lattice spacing $\epsilon$
merely plays the role of a bookkeeping device.
We now assume that
\begin{equation}
\exists\; \delta,\; C>0,\quad
\cases
{
\lim_{\epsilon\rightarrow0}\
{1\over\epsilon^{1+\delta}}\ |a^1_{\hat{\rm o}}|<C,&\cr
\hfil&\hfil\cr
\lim_{\epsilon\rightarrow0}\
{1\over\epsilon^{\delta}}\ |{\rm Re}\ E-{\rm Re}\ E_{\rm f}|<C,&\cr
\hfil&\hfil\cr
\lim_{\epsilon\rightarrow0}\
{1\over\epsilon^{\delta}}\ |{\rm Im}\ E-{\rm Im}\ E_{\rm f}|<C,&\cr
\hfil&\hfil\cr
\lim_{\epsilon\rightarrow0}\
{1\over\epsilon^{\delta}}\ |X|<C,&\cr
}
\label{MFparaepsilon}
\end{equation}
so that the formal expansion in powers of $\epsilon$
of the mean-field excitation
energies around any of the four points within the Brillouin Zone at which
the flux spectrum is gapless, yields the relativistic spectra
\begin{equation}
\label{massivespinondispersion}
|\vec\xi_{\vec k_l}|=
{\rm v_{f}}
\sqrt
{
\vec p^{\ 2}+m^2{\rm v_{f}}^2
}+O(\epsilon^2).
\end{equation}
The mean-field parameters of Eq. (\ref{massivespinondispersion})
are taken in the flux phase and
\begin{equation}
{\rm v_{f}}\;=\;J_2\;|{\rm Re}\ E|\ \epsilon,\quad
m\;\equiv \;
{J_1\;|X|\over2{\rm v_{f}}^2}.
\end{equation}

This simplification suggests that the
problem becomes tractable if we can argue that
the deviations of $a^1_{\hat{\rm o}}$ and $E$
from their flux phase values are negligible compared to
the deviations of the mean-field parameter $X$ from its flux phase value.
By inspection of the saddle-point solutions, one extracts that
on approaching the flux phase from below
(see  Fig.~\ref{scaling})
\begin{eqnarray}
&&
|a^1_{\hat{\rm o}}|\propto
\left(1-{(J_2/J_1)\over (J_2/J_1)_{\rm cr}}\right)^{\beta_1},
\quad \beta_1=1.52,
\nonumber\\
&&
|{\rm Re}\ E-{\rm Re}\ E_{\rm f}|\propto
\left(1-{(J_2/J_1)\over (J_2/J_1)_{\rm cr}}\right)^{\beta_2},
\quad \beta_2=1.79,
\nonumber\\
&&
|{\rm Im}\ E-{\rm Im}\ E_{\rm f}|\propto
\left(1-{(J_2/J_1)\over (J_2/J_1)_{\rm cr}}\right)^{\beta_3},
\quad \beta_3=1.77,
\nonumber\\
&&
|X|\propto
\left(1-{(J_2/J_1)\over (J_2/J_1)_{\rm cr}}\right)^{\beta_4},
\quad \beta_4=0.84.
\label{MFscaling}
\end{eqnarray}
The important insight that follows from
Eq.~(\ref{MFscaling})
is that $\beta_4={\rm min}\{\beta_1,\cdots,\beta_4\}$.
Thus, the deviation of $X$ from its flux value is more important than
the deviations of
$a^1_{\hat{\rm o}}$,
${\rm Re}\; E$ and
${\rm Im}\; E$ from their flux values.
We will now show how to recover the massive relativistic dispersion relation
of Eq.~(\ref{massivespinondispersion})
from the relativistic theory that was constructed in the previous sections
when $a^1_{\hat{\rm o}}$ and $J_2E$ are
assumed to take their flux phase values whereas $J_1X$ is non-vanishing.

In the flux gauge of appendix \ref{sec:flux}
the sixteen degrees of freedom of
Eq.\ (\ref{thecalN})
and
Eq.\ (\ref{thecalABCD})
have the mean-field values
\begin{eqnarray}
&&
\bar A^{eo}_{\mu}\ =\
\bar B^{eo}_{\mu}\ =\
\bar C^{eo}_{\mu}\ =\
\bar D^{eo}_{\mu}\ =\
0,\quad \mu=0,1,2,
\nonumber\\
&&
\bar N^{eo   }_n\ =\
{2\ X\over\bar\epsilon}\;
V^{  \  }_n,
\quad n=1,\dots,4,
\end{eqnarray}
where
\begin{eqnarray}
&&
V_1\;\equiv \ {1\over\sqrt{2}}\;({\rm i}\sigma^0-W_-)\;=\;
+{\rm i}e^{+{\rm i}{\pi\over4}W_-},
\nonumber\\
&&
V_2\;\equiv \ {1\over\sqrt{2}}\;(            W_+-W_3)\;=\;
+{\rm i}e^{-{\rm i}{\pi\over2}{1\over\sqrt{2}}(W_+-W_3)},
\nonumber\\
&&
V_3\;\equiv \ {1\over\sqrt{2}}\;(            W_++W_3)\;=\;
+{\rm i}e^{-{\rm i}{\pi\over2}{1\over\sqrt{2}}(W_++W_3)},
\nonumber\\
&&
V_4\;\equiv \ {1\over\sqrt{2}}\;({\rm i}\sigma^0+W_-)\;=\;
+{\rm i}e^{-{\rm i}{\pi\over4}W_-}.
\end{eqnarray}
Consequently, at mean-field
Eq.\ (\ref{Linfluxphase})
reduces to
\begin{equation}
\bar{\cal L}\;=\;
\bar{\cal L}_0\;+\;\bar{\cal L}_1\;+\;\bar{\cal L}_2.
\label{meanfieldlagrangianbelowflux}
\end{equation}

Here, if
\begin{equation}
q\;\equiv \;
\pmatrix
{
\Psi\cr
\Theta \cr
}\;=\;
\pmatrix
{
u\cr
w\cr
v\cr
z\cr
},
\end{equation}
then
\begin{equation}
\bar{\cal L}_0\;=\;
{\rm v_f}\; \bar q\;
{\rm i}\partial_{\mu}\;
q,
\end{equation}
\begin{equation}
\bar{\cal L}_1\;=\;
-{J_1\over16}\ \sum_{n=1}^4\
{\rm tr}\
\left[
\bar{\cal N}^{\   }_n\ \bar{\cal N}^{\dag}_{n}
\right]
+
{\rm v_f}\; \bar q
\; m{\rm v_f}\;
\bar{\sf M}\;
q,
\label{quarkmassterm}
\end{equation}
\begin{equation}
\bar{\cal L}_2\;=\;
0.
\end{equation}
The matrix $m\bar{\sf M}$ plays the role of a mass matrix for the $q$'s.
It is given in a 4$\times$4 dimensional representation by
\begin{equation}
\bar{\sf M}\;=\;
{{\rm i}\over\sqrt{2}}\;
\pmatrix
{
0&+V^{\   }_1&0&+V^{\   }_2\cr
-V^{\dag}_1&0&-V^{\dag}_3&0\cr
0&+V^{\   }_3&0&+V^{\   }_4\cr
-V^{\dag}_2&0&-V^{\dag}_4&0\cr
}.
\end{equation}
It can be checked that the matrix $\bar{\sf M}$,
which, in its full glory, is a 16 by 16 matrix,
has the alternating
eigenvalues $\pm1$. Hence, the dispersion relation
Eq.\ (\ref{massivespinondispersion})
is reproduced by
Eq.\ (\ref{meanfieldlagrangianbelowflux}).
The global SU(2)$\times$SU(2) color symmetry is broken by
Eq.\ (\ref{meanfieldlagrangianbelowflux})
all the way down to ${\cal Z}_2$.
Again, the generation of the masses for the four Dirac species
$u$, $w$, $v$ and $z$ is equivalent to the breaking of all the
continuous color symmetries.

Finally, by using a linear parametrization of the
scalar gauge fields fluctuations,
the non-linear parametrization of the gauge
fluctuations on the next-nearest-neighbor links of Eq.
(\ref{nonlinearparametrization}) and a linear parametrization of the
nearest-neighbor links fluctuations, we find from
Eq. (\ref{Linfluxphase}) and Eq. (\ref{quarkmassterm})
the intermediate effective theory
\begin{equation}
{\cal L}\;=\;{\cal L}_0 + \bar{\cal L}_1 + {\cal L}_1 + {\cal L}_2.
\end{equation}

\subsection{Symmetries of the intermediate effective theory in the flux phase.}
\label{subsec:symfluxcont}

In the past subsections we showed that it is possible to construct an
effective continuum theory for the slow modes of the system near the
flux regime. The effective theory that we constructed has all the right
ingredients for the scenario of section \ref{sec:Deconfinement}
to work. In particular,
below the flux phase, all the continuous gauge symmetries are broken
down to a discrete ${\cal Z}_2$ and the fermions are all massive. A
loop expansion
in the gauge coupling constant is well defined order by order and
yields a local action for the gauge fields. The only low energy degrees
of freedom left are the adjoint bosonic matter fields (which have
acquired dynamics through the fermions) and the gauge fields. Hence, the
effective action for these degrees of freedom will have the
form indicated in section \ref{sec:Deconfinement}.
We will present elsewhere the calculation
of this effective bosonic action \cite{Mudry}.
By construction, it will have the two crucial features that
all continuous symmetries are broken and that the only remaining
symmetry is discrete namely ${\cal Z}_2$.
In view of the arguments given
at the beginning of section \ref{sec:Deconfinement},
the spinon quantum numbers
are present in the spectrum of the theory provided the effective scales for
the kinetic energies of the gauge and matter fields are large enough.
By varying the ratio $J_2/J_1$, the parameters
$\beta$ and $\kappa$ in Eq.~(\ref{fradkinaction}) of the effective
theory change. However, the mean-field results alone do not guarantee
that it is possible to move from the
confining regime to the Anderson-Higgs phases of the effective theory.
If this can be done, then we have a mechanism which does not
break parity and
time-reversal and allows for the presence of the color spinon quantum
number in the spectrum of the theory. The only other mechanism
that we are aware of has its origin in the chiral spin liquid
for which the effective continuum limit contains a Chern-Simons term
\cite{Wen 1989}. The presence of the Chern-Simons term
in the bosonized action removes the interactions responsible otherwise
for the confinement of the fundamental color charge \cite{Fradkin2 1991}.

Let us take the limit $J_2/J_1\rightarrow\infty$ for which
the Heisenberg model reduces to
two independent, nearest-neighbor antiferromagnetic Heisenberg models defined
on the even and odd sublattices, respectively.
Although the exact ground state of the nearest-neighbor, spin-1/2,
antiferromagnetic Heisenberg model is not known,
there are strong evidences that it carries
N\' eel order \cite{Manousakis 1991}.
We therefore expect to find
degenerate ground states carrying independent N\' eel order on the even
and odd sublattices and our mean-field theory must be unstable
\cite{digressiononBZA} to the N\' eel ordering.
The natural question to ask is whether Eq. (\ref{continummlimitJ1=0})
can account for this instability or if we need to include fluctuations
with higher energy and/or shorter wavelength. Since N\' eel ordering
requires the breaking of a global symmetry, let us see what symmetries
beside the local gauge symmetry Eq. (\ref{continummlimitJ1=0})
possesses.

When $J_2/J_1\rightarrow\infty$,
there exists an additional global U(4) (flavor) symmetry
under the transformation
\begin{equation}
\pmatrix
{
u\cr
w\cr
v\cr
z\cr
}
\rightarrow
V\;
\pmatrix
{
u\cr
w\cr
v\cr
z\cr
},
\quad
\forall V\in{\rm U(4)}={\rm U(1)}\times{\rm SU(4)}.
\end{equation}
A possible manifestation of an instability of the flux phase is the
dynamical breaking of the global flavor symmetry
as a result of the gauge fluctuations.
We have listed in table \ref{table1} of appendix \ref{sec:stagmag}
all 15 lattice Hamiltonians $H_a$ expressed in terms of the spinon operators
defined by Eq. (\ref{spininspinonbasis}) which yield all
15 generators $\bar q\ T_a\ q$ breaking the SU(4) flavor symmetry.
One sees that the 9 generators $T_1$, $T_3$, $T_4$, $T_6$, $T_7$,
$T_9$, $T_{11}$, $T_{13}$ and $T_{15}$ are all related to different types
of valence bond ordering which do not break the spin symmetry
\cite{dangerousirrelevance}.
The remaining 6 generators are related to spin densities and spin currents.
For example, $\bar q\ T_2\ q$, measures the N\' eel ordering on the
unit cell of the even sites whereas $\bar q\ T_4\ q$ measures
the N\' eel ordering on the odd sites.
Finally, the generator of U(1) is also related to valence bond ordering.
We thus conclude that Eq. (\ref{continummlimitJ1=0}) has enough symmetries
to account for the instability towards N\' eel ordering. In fact our
effective action could be unstable to many other types of ordering
including Peierls ordering.

It is an open problem to show from
Eq. (\ref{continummlimitJ1=0})
that the flavor symmetry is broken dynamically
\cite{Diamantini 1993}.
The difficulty comes from the non-perturbative
nature of such a mechanism.
For example, one can formally integrate the fermions in
Eq. (\ref{continummlimitJ1=0}). To one loop, the resulting action
for the gauge fields is finite but non-local
\cite{Jackiw 1981}. However, higher loop contributions
are infrared divergent due to the masslessness of the fermions.
The infrared problem can be bypassed if one chooses
a dimensionless expansion parameter such as $1/{\rm N_F}$
with ${\rm N_F}$ the flavor number,
instead of the dimensionfull gauge coupling.
Within this approximation, the lowest order in $1/{\rm N_F}$
already predicts spontaneous symmetry breaking of SU(4)
\cite{Appelquist 1990} although this result has to be taken with caution
\cite{Diamantini 1993}.
When $(J_2/J_1)_{\rm cr}<J_2/J_1<\infty$, the effective theory
Eq. (\ref{Linfluxphase})
has only a U(1) global flavor symmetry.
For a small number of flavors (such as $4=2\times2$ in our case),
it is likely that the instability
of the flux phase with respect to the quantum fluctuations at
$(J_2/J_1)_{\rm cr}$ is preempted by a first order transition
from a state carrying long range spin-spin correlations to a spin liquid.
In any case, Eq. (\ref{Linfluxphase}) seems rich enough
to describe the phase diagram of the $J_1-J_2$ spin-1/2 antiferromagnetic
Heisenberg model
as inferred from a bosonic representation of a Sp(N)
generalization of the model \cite{Sachdev 1991}.

\section{The one dimensional case.}
\label{sec:onedim}

In the previous section we have discussed a scenario for separation of
spin and charge (or the occurence of unusual quantum numbers for
excitations) to take place in two dimensional systems.
In this section we want to see what picture this scenario yields
about the one dimensional Heisenberg chain.
Our purpose is twofold. Firstly, the
one dimensional Heisenberg chain is known to exhibit separation of spin
and charge in the sense that it has in its excitation spectrum gapless
excitations with spin-1/2. These excitations are nowadays referred
to as the spinons of the one dimensional chain. Secondly, we want
to make a comparison with the two dimensional problem.
In particular, we want to see what is the relation, if any,
between the gapless spinons of the one dimensional chain and the
spinons of the gauge picture.

The one dimensional Heisenberg chain has
been extensively studied by many methods and much is known about the
physics of this system.
The Bethe Ansatz yields the ground state and the excitations above it.
Conformal field theory gives the critical exponents of the correlation
functions. The excitation spectrum thus found consists of gapless
integer spin excitations (``spin waves'') and
spin-1/2 gapless excitations (``spinons'').

The integer spin excitations are the expected spin flips.
The gapless spin-1/2 excitations are the
ones we are interested in. The simplest way to see what the ``spinons''
are is by means of the Jordan-Wigner \cite{Jordan 1928}
transformation of the spin-1/2 operators:
\begin{eqnarray}
&&
c^{\   }_i\;\equiv\;
e^
{
+{\rm i}\pi\sum_{j<i}\ S^+_j\ S^-_j
}
\;
S^-_i,
\nonumber\\
&&
c^{\dag}_i\;\equiv\;
S^+_i\
e^
{
-{\rm i}\pi\sum_{j<i}\ S^+_j\ S^-_j
}.
\label{spinless}
\end{eqnarray}
{}From Eq. (\ref{spininpsibasis}),
we recognize in the raising and lowering spin-1/2 operators $S^{\pm}_i$
the color singlet ``baryon'' operators $b^{\dag}_i$ and $b^{\   }_i$,
respectively:
\begin{eqnarray}
&&
S^+_i\equiv S^1_i+{\rm i}S^2_i\;=\;b^{\dag}_i,
\nonumber\\
&&
S^-_i\equiv S^1_i-{\rm i}S^2_i\;=\;b^{\   }_i,
\label{rvslspinoperator}
\end{eqnarray}
whereas the ``mesonic'' operator $m^{\dag}_i$ is related to
\begin{equation}
S^+_i\;S^-_i\;=\;
{1\over2}\;+\;S^3_i\;=\;{1\over2}m^{\   }_i.
\end{equation}
The fermion operators $c_i$ defined by the Jordan-Wigner transformation
can be viewed as spinless fermions in the sense that they do not carry an
explicit spin index. Naturally, the expression of Eq. (\ref{spinless})
depends on the choice of the spin orientation. In this sense the $c$'s do
transform under the global SU(2) of spin. But, and this is what matters
for a comparison with the gauge picture, they are singlets under the
local SU(2) (color) and consequently gauge invariant. Thus, these are allowed
excitations which are compatible with confinement of the color
degrees of freedom. Another important feature of the ``spinon'' operators
of Eq. (\ref{spinless}) is the fact that they disturb the boundary
conditions and, hence, are topological excitations. These operators are
non-local and gauge invariant. In contrast, the spinons of the gauge
picture are local but not invariant under local SU(2) (color) transformations.
In addition, the spinons of the gauge picture, are absent from the
physical spectrum if the system is realized in a confining state.
Clearly, the Jordan-Wigner spinons are not.

To sum up, the fermion $c_i$, when expressed in terms of the
baryons $b_i$ and mesons $m_i$, give a non-local and non-linear
realization of an operator transforming like the fundamental
representation of SU(2) under spin rotations and transforming
like a color singlet under gauge transformations.
This is very suggestive of how confinement precludes any simple
description of the spin-1/2 low energy sector of the one dimensional
Heisenberg model in terms of color carrying spinons.

Let us see what the SU(2) gauge approach tells us about the one
dimensional case. In particular we would like to see how the known
physics of this system is recovered in the gauge field approach.
It is straightforward to apply
the methods of the previous sections to the one dimensional case. In a
separate publication we do so \cite{Mudry}. Here, we will discuss several
qualitative differences that arise when this picture is applied to
the one dimensional case.

Along the lines of section \ref{subsec:MFd>2}, we try the
mean-field Ansatz
\begin{eqnarray}
\label{Ansatzd=1}
&&
\bar A^{\ }_{\hat{\rm o}it}=A^{\ }_{\hat{\rm o}},
\\
&&
\bar W^{\   }_{ijt}=
\cases{
-X\sigma^3& if $j=i+1$,\cr
-{\rm Re}\ E\ \sigma^1+{\rm Im}\ E\ \sigma^2& if $j=i+2$, $i$ even,\cr
-{\rm Re}\ E\ \sigma^1-{\rm Im}\ E\ \sigma^2& if $j=i+2$, $i$ odd,\cr
0&otherwise.\cr
}
\nonumber
\end{eqnarray}
It describes a spin liquid if and only if
\begin{equation}
A_{\hat{\rm o}}\;=\;{1\over2}\;a^1_{\hat{\rm o}}\;\sigma^1.
\end{equation}
This Ansatz satisfies all of the requirements: a) it is translationally
invariant (up to gauge transformations),
b) it respects all point group symmetries and time reversal invariance
(up to gauge transformations) and c) it achieves a
complete breaking of the local continuous gauge symmetry through the
operators
\begin{eqnarray}
&&
P^{{\rm e}}_{i  }=
W^{\   }_{i(i+1)}
W^{\   }_{(i+1)(i+2)}
W^{\dag}_{i(i+2)},
\\
&&
P^{{\rm o}}_{i+1}=
W^{\   }_{(i+1)(i+2)}
W^{\   }_{(i+2)(i+3)}
W^{\dag}_{(i+1)(i+3)},
\quad\forall i\;{\rm even}.
\nonumber
\end{eqnarray}
In particular, just as in the two dimensional case,
it leaves a discrete ${\cal Z}_2$ symmetry unbroken.

However, discrete gauge symmetries play a very different role in
one and two space dimensions. This is so since the lower (space-time)
critical dimension for gauge theories with a {\it discrete} symmetry
group (such as ${\cal Z}_2$ in the case of interest for our problem) is
three. Hence, in the case of the one dimensional
quantum antiferromagnet, which lives in two space-time dimensions, the
effective ${\cal Z}_2$ gauge theory along the topmost line of the
phase diagram in Fig.~\ref{phasediagram}
is below the lower critical dimension and thus is
always in a confining phase. No phase transition is
encountered along the line
$(\beta,\kappa,\lambda)=(\beta,\infty,\infty)$ and there can only
be one phase in the phase diagram, the confinement-Higgs phase.
In this phase, the spinon carrying the fundamental color charge is
confined and is not part of the spectrum of finite energy states.
Clearly, the spin-1/2 excitations of the one dimensional
quantum Heisenberg antiferromagnet {\it are not} the spinons of the gauge
picture, as has been claimed rather loosely in the literature of the subject.
As we stressed above, the fermion states described by the Jordan-Wigner
operators are manifestly gauge invariant under the local SU(2) symmetry
and are unaffected by confinement.

\section{Conclusion.}
\label{sec:concl}

We have studied the issue of spin and charge separation for
strongly correlated electronic systems in two spatial dimensions by studying
the frustrated Heisenberg model for spin-1/2 in a representation of the
spin degrees of freedom in terms of fermions (spinons) and gauge fields.
Our treatment preserves explicitly the existing SU(2) local gauge symmetry
which allows us to treat on the same footing the Affleck-Marston
and the Anderson order parameters.
We have given a set of sufficient
conditions for the system to have excitations
in which the spinons carry the separated spin quantum numbers, i.e.,
spin-1/2 degrees of freedom, {\it inspite of the existence of strong
gauge fluctuations}.

We solved the saddle-point equations
for an isotropic mean-field Ansatz first proposed by Wen and
depending on four real parameters. There exists a second
order phase transition to a new flux phase which is triggered by the
competition between the nearest $J_1$ and next-nearest-neighbor $J_2$
antiferromagnetic couplings of the model.
The flux phase is chosen for strong enough $J_2$.
Below the flux phase, a gap for the spinons opens and reaches a maxima for
$J_2/J_1\approx 0.7$.
Although no level crossing occurs for the mean-field ground states below
the flux phase, there is a level crossing for the excited states
when $J_2/J_1\approx 0.7$. On approaching the pure nearest-neighbor limit,
the features of a tight-binding Fermi surface at half-filling emerge.
The phase $0<J_2/J_1<(J_2/J_1)_{{\rm cr}}$, the s-RVB phase,
is characterized by an energy gap and the breaking of all continuous
gauge symmetries.

We derived a continuum theory for the soft degrees of freedom in
the vicinity of the flux phase which describes bosonic gauge and matter
fields interacting with the spinons.
In the flux phase, the continuum theory
is likely to be unstable towards formation of long range antiferromagnetic
order or dimer ordering. A more detailed calculation is needed to
decide which ordering is chosen by the system but, in any case, deconfinement
of the spinons is not expected to take place.
We showed that below the flux phase, the only important remaining
soft modes are those of a ${\cal Z}_2$ gauge theory in three space-time
dimensions due to the breaking of all continuous gauge symmetries
by the s-RVB Ansatz.
Deconfinement of the gauge spinons is then allowed in the phase
diagram of a generic theory with the same symmetry attributes but
in which the gauge and matter fields have independent kinetic energy scales.
In our case, the calculation of the ratio of the {\it effective} energy
scales for the gauge and matter fields through the integration of the spinons
is needed to decide whether the system is in the confining or deconfining
regime. {\it Consequently, although deconfinement of the spinons is allowed
it is not guaranteed} since it depends on the values of microscopic
parameters. This result stands in sharp contrast
to the analysis of the one dimensional quantum antiferromagnet which
predicts that {\it the gauge spinons are always confined},
irrespectively of the value of the effective energy scales for the gauge
and matter fields, although the system still exhibits separation of spin and
charge. The mechanism for separation of spin and charge in one dimension
is topological and it is unrelated to the issue of confinement.

\section*{\acknowledgements}

We have benefitted from useful discussions with J.-F. Laga\" e.
This work was supported in part by NSF grants No.~ DMR91-22385 at the
Department of Physics of the University of Illinois at Urbana-Champaign,
DMR89-20538 at the Materials Research Laboratory of the University of
Illinois, an IBM Graduate Student Fellowship and by a grant of the
Research Board of the University of Illinois at Urbana-Champaign.
We have made extensive use of the
facilities of the Materials Research Laboratory Center for Computations.

\appendix
\section{The flux phase}
\label{sec:flux}

We have found in section \ref{sec:s-RVBMF}
that above a critical value of the coupling constant $J_2/J_1$, the
s-RVB phase collapses to the case
\begin{equation}
\vec a_{\hat{\rm o}}\;=\;0,\quad
{\rm Re}\ E\;=\;{\rm Im}\ E,\quad
X\;=\;0,
\end{equation}
i.e., the spectrum takes the simple form
\begin{equation}
|\vec\xi_{\vec k}|\;=\;
J_2\;|{\rm Re}\ E|\;
\sqrt{\cos^2k_x\cos^2k_y\;+\;\sin^2k_x\sin^2k_y}.
\end{equation}
The spectrum is thus gapless and linearization around the four minima
of Eq.\ (\ref{fluxminima})
yields a massless relativistic dispersion relation.

Another characterization of this phase can be made. We introduce the
orthogonal and normalized basis
\begin{equation}
W_-=-{\sigma^1+\sigma^2\over\sqrt{2}},\quad
W_+=-{\sigma^1-\sigma^2\over\sqrt{2}},\quad
W_3=-\sigma^3,
\end{equation}
of the Lie Algebra su(2). This basis satisfies the su(2) Algebra with
\begin{equation}
W_-W_+={\rm i}W_3,\quad
W_+W_3={\rm i}W_-,\quad
W_3W_-={\rm i}W_+,
\end{equation}
and can be used to rewrite
\begin{equation}
\bar W^{\   }_{ijt}=
\cases
{
0&along the links $\hat x$ and $\hat y$,\cr
|E|W_+&along the link $\hat x_+$,\cr
|E|W_-&along the link $\hat x_-$,\cr
}
\end{equation}
whenever $J_2$ is larger than its critical value. This regime
is characterized uniquely by the most elementary path ordered
operator along a closed path:
\begin{equation}
\bar P_{\diamond}\;\equiv \;
W^{\   }_-
W^{\   }_+
W^{\dag}_-
W^{\dag}_+=-\sigma^0.
\label{diamondpath}
\end{equation}
Here, $\sigma^0$ is the neutral element of SU(2).
Eq.\ (\ref{diamondpath}) tells us that there is a mean-field flux
of $\pi$ per elementary plaquette, a situation commonly referred to as the
flux phase \cite{Kotliar 1988,Affleck 1988}.

As of now, the mean-field Ansatz had an explicit translational invariance.
In order to describe small fluctuations of the link and time-like su(2)
degrees of freedom around their mean-field value in the flux phase, it
turns out to be convenient to choose a gauge in which the color structure
on the next-nearest-neighbor links becomes trivial. The price to be
payed is a reduction of the translational symmetry. First we need some
notation.

We introduce eight sublattices
(see  Fig.~\ref{evenoddsublattices})
by partitioning first
the square lattice $\Lambda$,
whose sites $i\in {\cal Z}^2$ are labelled by pairs of integer,
into two interpenetrating sublattices
$\Lambda^{\ }_{\rm e}$ and $\Lambda^{\ }_{\rm o}$:
\begin{eqnarray}
&&
\Lambda^{\ }_{\rm e}\;=\;
\left\{\;
i\in\Lambda\;|\;i_1+i_2=0\ {\rm mod}\ 2\;
\right\},
\nonumber\\
&&
\Lambda^{\ }_{\rm o}\;=\;
\left\{\;
i\in\Lambda\;|\;i_1+i_2=1\ {\rm mod}\ 2\;
\right\}.
\end{eqnarray}

The even and odd sublattices are then partitioned into four sublattices
$\Lambda^{(a)}_{\rm e},\ \Lambda^{(a)}_{\rm o},\ a=1,2,3,4$ by
choosing
\begin{eqnarray}
&&
\Lambda^{(1)}_{\rm e}\;=\;
\left\{
i\in\Lambda^{\ }_{\rm e}\;|\;\exists\ j\in {\cal Z}^2, i=j_1\ 2\hat x_-+j_2\
2\hat
x_+
\right\},
\nonumber\\
&&
\Lambda^{(1)}_{\rm o}\;=\;
\left\{
i\in\Lambda^{\ }_{\rm o}\;|\;\exists\ j\in \Lambda^{(1)}_{\rm e}, i=j+\hat x
\right\},
\end{eqnarray}
and using translations to define
\begin{eqnarray}
&&
\Lambda^{(2)}_{\rm e}\;=\;
\left\{
i\in\Lambda^{\ }_{\rm e}\;|\;\exists\ j\in \Lambda^{(1)}_{\rm e}, i=j+\hat x_-
\right\},
\nonumber\\
&&
\Lambda^{(3)}_{\rm e}\;=\;
\left\{
i\in\Lambda^{\ }_{\rm e}\;|\;\exists\ j\in \Lambda^{(2)}_{\rm e}, i=j+\hat x_+
\right\},
\nonumber\\
&&
\Lambda^{(4)}_{\rm e}\;=\;
\left\{
i\in\Lambda^{\ }_{\rm e}\;|\;\exists\ j\in \Lambda^{(3)}_{\rm e}, i=j-\hat x_-
\right\},
\end{eqnarray}
and similarly for the odd sublattices
$\Lambda^{(2)}_{\rm o}$,
$\Lambda^{(3)}_{\rm o}$ and
$\Lambda^{(4)}_{\rm o}$.
{}From now on, $i$ will always be taken to belong to $\Lambda^{(1)}_{\rm e}$.

The fermionic doublets will be renamed
\begin{eqnarray}
&&
\psi^{\   }_{i}\rightarrow f^{e1}_{i},\qquad
\psi^{\   }_{i+\hat x}\rightarrow f^{o1}_{i},
\nonumber\\
&&
\psi^{\   }_{i+\hat x_-}\rightarrow f^{e2}_{i},\qquad
\psi^{\   }_{i+\hat x+\hat x_-}\rightarrow f^{o2}_{i},
\nonumber\\
&&
\psi^{\   }_{i+\hat x_-+\hat x_+}\rightarrow f^{e3}_{i},\qquad
\psi^{\   }_{i+\hat x+\hat x_-+\hat x_+}\rightarrow f^{o3}_{i},
\nonumber\\
&&
\psi^{\   }_{i+\hat x_+}\rightarrow f^{e4}_{i},\qquad
\psi^{\   }_{i+\hat x+\hat x_+}\rightarrow f^{o4}_{i}.
\end{eqnarray}
In the same way, there will be $2\times4=8$ independent time-like gauge
fields:
\begin{equation}
A^{\ }_{\hat{\rm o}i}\rightarrow
A^{e1}_{\hat{\rm o}i},
\cdots,
A^{\ }_{\hat{\rm o}(i+\hat x+\hat x_+)}\rightarrow
A^{o4}_{\hat{\rm o}i},
\end{equation}
$2\times8=16$ next-nearest-neighbor link degrees of freedom:
\begin{eqnarray}
\label{allnextnearestneighborlinks}
W^{\ }_{i(i+\hat x_-)}\rightarrow
Q^{e1}_{\hat x_-i},&&\cdots
\\
&&\cdots,
W^{\ }_{(i+\hat x+\hat x_+)(i+\hat x+2\hat x_+)}\rightarrow
Q^{o4}_{\hat x_+i},
\nonumber
\end{eqnarray}
and $2\times8=16$ nearest-neighbor link degrees of freedom:
\begin{eqnarray}
\label{allnearestneighborlinks}
W^{\ }_{i(i+\hat x)}\rightarrow
M^{e1}_{\hat xi},&&\cdots
\\
&&\cdots,
W^{\ }_{(i+\hat x+\hat x_+)(i+\hat x+\hat x_++\hat y)}\rightarrow
M^{o4}_{\hat yi}.
\nonumber
\end{eqnarray}

We are now ready to perform the staggered gauge transformation
\begin{eqnarray}
&&
f^{e1}_i\rightarrow {\rm sgn}(i)\ (+\sigma^0)\ f^{e1}_i,\quad\
f^{o1}_i\rightarrow {\rm sgn}(i)\ (+\sigma^0)\ f^{o1}_i,
\nonumber\\
&&
f^{e2}_i\rightarrow {\rm sgn}(i)\ (-{\rm i}W_-)f^{e2}_i,\quad
f^{o2}_i\rightarrow {\rm sgn}(i)\ (-{\rm i}W_-)f^{o2}_i,
\nonumber\\
&&
f^{e3}_i\rightarrow {\rm sgn}(i)\ (-{\rm i}W_3)f^{e3}_i,\quad
f^{o3}_i\rightarrow {\rm sgn}(i)\ (-{\rm i}W_3)f^{o3}_i,
\nonumber\\
&&
f^{e4}_i\rightarrow {\rm sgn}(i)\ (+{\rm i}W_+)f^{e4}_i,\quad
f^{o4}_i\rightarrow {\rm sgn}(i)\ (+{\rm i}W_+)f^{o4}_i,
\nonumber\\
&&
\end{eqnarray}
for all $i\in\Lambda^{(1)}_{\rm e}$.
Here, ${\rm sgn}(i)$ is the sign function
which distinguishes between even and odd sites of sublattice
$\Lambda^{(1)}_{\rm e}$.
Under this gauge transformation, the mean-field Ansatz for the
nearest-neighbor links takes the simple form
\begin{eqnarray}
\bar Q^{e1}_{\hat x_-i}=&&+{\rm i}|E|\;\sigma^0,\quad
\bar Q^{e1}_{\hat x_+i}=  -{\rm i}|E|\;\sigma^0,
\nonumber\\
\bar Q^{e2}_{\hat x_-i}=&&+{\rm i}|E|\;\sigma^0,\quad
\bar Q^{e2}_{\hat x_+i}=  +{\rm i}|E|\;\sigma^0,
\nonumber\\
\bar Q^{e3}_{\hat x_-i}=&&+{\rm i}|E|\;\sigma^0,\quad
\bar Q^{e3}_{\hat x_+i}=  +{\rm i}|E|\;\sigma^0,
\nonumber\\
\bar Q^{e4}_{\hat x_-i}=&&+{\rm i}|E|\;\sigma^0,\quad
\bar Q^{e4}_{\hat x_+i}=  -{\rm i}|E|\;\sigma^0,
\end{eqnarray}
and similarly on $\Lambda^{(1)}_{\rm o}$. One easily verifies that
Eq.\ (\ref{diamondpath})
still holds.

\section{Continuum limit for some lattice Hamiltonians.}
\label{sec:stagmag}

We have collected in this appendix all the 15 lattice Hamiltonians
which yield the 15 generators of the global SU(4) symmetry of the
Lagrangian density Eq. (\ref{continummlimitJ1=0}).
They are displayed in table \ref{table1} below.

\vfill\eject

\begin{figure}
\caption{
Phase diagram for a gauge-Higgs lattice theory in 2+1 dimensional space-time.
$\beta$ is the energy scale for the gauge fields,
$\kappa$ that for the matter (Higgs). The amplitude of the Higgs is
frozen ($\lambda=\infty$). (a) shows the case of the matter (Higgs) in the
fundamental representation. (b) shows the case of the matter (Higgs) in the
adjoint representation.}
\label{phasediagram}
\end{figure}

\begin{figure}
\caption
{
The solutions $a_{\hat{\rm o}}^1$, ${\rm Re}\ E$, ${\rm Im}\ E$ and $X$
as a function of the dimensionless ratio $J_2/J_1$ of the saddle-point
equations for Wen's Ansatz.
}
\label{mfsolutions}
\end{figure}

\begin{figure}
\caption
{
The upper branch $+|\vec\xi_{\vec k}|$ of the MF single-particle
excitations (spinons) in the first quadrant of the Brillouin Zone for:
(a) $J_2/J_1=1.4$,
(b) $J_2/J_1=0.9$,
(c) $J_2/J_1=0.5$,
(d) $J_2/J_1=0$.
The minima move from the edges (a) and (b) to the diagonal (c) and (d).
}
\label{spinonenergysurface}
\end{figure}

\begin{figure}
\caption
{
Phase diagram of Wen's mean-field Ansatz. There exists a second order
phase transition into a phase, the flux phase, without any gap
to the mean-field excitations. Below the threshold
$(J_2/J_1)_{\rm cr}\approx 1.342$, the mean-field
excitations have a gap, except at the BZA point $J_2/J_1=0$.
The critical point $(J_2/J_1)_{\rm cr}$ is infrared unstable
as indicated by the RG flow of $J_2/J_1$.
The location of the minima in the s-RVB phase,
$0<J_2/J_1<(J_2/J_1)_{\rm cr}$, jumps discontinuously
for $J_2/J_1\approx 0.7$, signaling a level crossing of the excitations.
The values of the gap are plotted in units of $J_1$.
}
\label{MFRG}
\end{figure}

\begin{figure}
\caption
{
The scaling exponents $\beta_1$, $\beta_2$, $\beta_3$ and $\beta_4$
of the s-RVB parameters
$a_{\hat{\rm o}}^1$, ${\rm Re}\ E$, ${\rm Im}\ E$ and $X$
as a function of $J_2/J_1$ can be read from a log-log plot.
We have used $(J_2/J_1)_{\rm cr}=1.342$ and
${\rm Re}\ E_{\rm f}={\rm Im}\ E_{\rm f}=0.339$.
}
\label{scaling}
\end{figure}

\begin{figure}
\caption
{
Partitioning of the square lattice $\Lambda$ into eight sublattices.
A sublattice is labelled by the letter $e$ or $o$ for even or odd
respectively and by an integer between 1 and 4.
}
\label{evenoddsublattices}
\end{figure}

\vfill\eject\widetext

\begin{table}
\squeezetable
\caption
{
The 15 bilinears $\int d^2x\ \bar q\ T_a\ q$
are given together with their lattice counterparts.
Here, $T_a$, $a=1,\cdots,15$ are the generators of SU(4) flavor.
The spinons $s$ were introduced in
section II
and are relabelled according to the lattice partitioning of appendix A.
The $u$, $w$, $v$ and $z$ are defined in section V.
}
\begin{tabular}{llr}
$a$
&
{\it Lattice Hamiltonian} $H_a$
&
$\int d^2x\ \bar q\ T_a\ q$
\\ \hline
\hfill&\hfill&\hfill\\1&
$
-{\rm i}\sum_{i\in\Lambda^e_1}
\big[
s^{e1\dag}_i\sigma^0s^{e2\   }_i+
s^{e4\dag}_i\sigma^0s^{e3\   }_i-
{\rm H.C.}
\big]
$
&
$+\int\ d^2x\left[\bar uv+\bar vu\right]$
\\
\hfill&\hfill&\hfill\\2&
$
+\sum_{i\in\Lambda^e_1}
\big[
s^{e1\dag}_i\sigma^3s^{e1\   }_i-
s^{e2\dag}_i\sigma^3s^{e2\   }_i+
s^{e3\dag}_i\sigma^3s^{e3\   }_i-
s^{e4\dag}_i\sigma^3s^{e4\   }_i+
{\rm H.C.}
\big]
$
&
$-{\rm i}\int\ d^2x\left[\bar uv-\bar vu\right]$
\\
\hfill&\hfill&\hfill\\3&
$
-{\rm i}\sum_{i\in\Lambda^e_1}
\big[
-s^{e1\dag}_i\sigma^0s^{e4\   }_i
+s^{e2\dag}_i\sigma^0s^{e3\   }_i
-{\rm H.C.}
\big]
$
&
$+\int\ d^2x\left[\bar uu-\bar vv\right]$
\\
\hfill&\hfill&\hfill\\4&
$
-{\rm i}\sum_{i\in\Lambda^o_1}
\big[
 s^{o1\dag}_i\sigma^0s^{o2\   }_i
+s^{o4\dag}_i\sigma^0s^{o3\   }_i
-{\rm H.C.}
\big]
$
&
$+\int\ d^2x\left[\bar wz+\bar zw\right]$
\\
\hfill&\hfill&\hfill\\5&
$
+\sum_{i\in\Lambda^o_1}
\big[
s^{o1\dag}_i\sigma^3s^{o1\   }_i-
s^{o2\dag}_i\sigma^3s^{o2\   }_i+
s^{o3\dag}_i\sigma^3s^{o3\   }_i-
s^{o4\dag}_i\sigma^3s^{o4\   }_i+
{\rm H.C.}
\big]
$
&
$-{\rm i}\int\ d^2x\left[\bar wz-\bar zw\right]$
\\
\hfill&\hfill&\hfill\\6&
$
-{\rm i}\sum_{i\in\Lambda^o_1}
\big[
-s^{o1\dag}_i\sigma^0s^{o4\   }_i
+s^{o2\dag}_i\sigma^0s^{o3\   }_i
-{\rm H.C.}
\big]
$
&
$+\int\ d^2x\left[\bar ww-\bar zz\right]$
\\
\hfill&\hfill&\hfill\\7&
$
-{\rm i}\sum_{i\in\Lambda^e_1}
\big[
 s^{e1\dag}_i\sigma^0s^{o3\   }_{i-2\hat x_+-2\hat x_-}
-s^{e1\dag}_i\sigma^0s^{o4\   }_{i-2\hat x_+}
+s^{e2\dag}_i\sigma^0s^{o3\   }_{i-2\hat x_+}
-s^{e2\dag}_i\sigma^0s^{o4\   }_{i-2\hat x_+}
$
\hfill&\hfill\\
\hfill&\hfill
$
-s^{e3\dag}_i\sigma^0s^{o1\   }_{i}
-s^{e3\dag}_i\sigma^0s^{o2\   }_{i}
+s^{e4\dag}_i\sigma^0s^{o1\   }_{i}
+s^{e4\dag}_i\sigma^0s^{o2\   }_{i-2\hat x_-}
-{\rm H.C.}
\big]
$
&
$+\int\ d^2x\left[\bar uw+\bar wu\right]$
\\
\hfill&\hfill&\hfill\\8&
$
-\sum_{i\in\Lambda^e_1}
\big[
 s^{e1\dag}_i\sigma^3s^{o3\   }_{i-2\hat x_+-2\hat x_-}
-s^{e1\dag}_i\sigma^3s^{o4\   }_{i-2\hat x_+}
+s^{e2\dag}_i\sigma^3s^{o3\   }_{i-2\hat x_+}
-s^{e2\dag}_i\sigma^3s^{o4\   }_{i-2\hat x_+}
$
\hfill&\hfill\\
\hfill&\hfill
$
-s^{e3\dag}_i\sigma^3s^{o1\   }_{i}
-s^{e3\dag}_i\sigma^3s^{o2\   }_{i}
+s^{e4\dag}_i\sigma^3s^{o1\   }_{i}
+s^{e4\dag}_i\sigma^3s^{o2\   }_{i-2\hat x_-}
+{\rm H.C.}
\big]
$
&
$-{\rm i}\int\ d^2x\left[\bar uw-\bar wu\right]$
\\
\hfill&\hfill&\hfill\\9&
$
-{\rm i}\sum_{i\in\Lambda^e_1}
\big[
 s^{e1\dag}_i\sigma^0s^{o2\   }_{i-2\hat x_-}
-s^{e1\dag}_i\sigma^0s^{o1\   }_{i}
+s^{e2\dag}_i\sigma^0s^{o2\   }_{i}
-s^{e2\dag}_i\sigma^0s^{o1\   }_{i}
$
\hfill&\hfill\\
\hfill&\hfill
$
-s^{e3\dag}_i\sigma^0s^{o3\   }_{i}
-s^{e3\dag}_i\sigma^0s^{o4\   }_{i}
+s^{e4\dag}_i\sigma^0s^{o3\   }_{i-2\hat x_-}
+s^{e4\dag}_i\sigma^0s^{o4\   }_{i}
-{\rm H.C.}
\big]
$
&
$+\int\ d^2x\left[\bar uz+\bar zu\right]$
\\
\hfill&\hfill&\hfill\\10&
$
-\sum_{i\in\Lambda^e_1}
\big[
 s^{e1\dag}_i\sigma^3s^{o2\   }_{i-2\hat x_-}
-s^{e1\dag}_i\sigma^3s^{o1\   }_{i}
+s^{e2\dag}_i\sigma^3s^{o2\   }_{i}
-s^{e2\dag}_i\sigma^3s^{o1\   }_{i}
$
\hfill&\hfill\\
\hfill&\hfill
$
-s^{e3\dag}_i\sigma^3s^{o3\   }_{i}
-s^{e3\dag}_i\sigma^3s^{o4\   }_{i}
+s^{e4\dag}_i\sigma^3s^{o3\   }_{i-2\hat x_-}
+s^{e4\dag}_i\sigma^3s^{o4\   }_{i}
+{\rm H.C.}
\big]
$
&
$-{\rm i}\int\ d^2x\left[\bar uz-\bar zu\right]$
\\
\hfill&\hfill&\hfill\\11&
$
-{\rm i}\sum_{i\in\Lambda^e_1}
\big[
 s^{e1\dag}_i\sigma^0s^{o1\   }_{i}
+s^{e1\dag}_i\sigma^0s^{o2\   }_{i-2\hat x_-}
-s^{e2\dag}_i\sigma^0s^{o1\   }_{i}
-s^{e2\dag}_i\sigma^0s^{o2\   }_{i}
$
\hfill&\hfill\\
\hfill&\hfill
$
+s^{e3\dag}_i\sigma^0s^{o3\   }_{i}
-s^{e3\dag}_i\sigma^0s^{o4\   }_{i}
+s^{e4\dag}_i\sigma^0s^{o3\   }_{i-2\hat x_-}
-s^{e4\dag}_i\sigma^0s^{o4\   }_{i}
-{\rm H.C.}
\big]
$
&
$+\int\ d^2x\left[\bar wv+\bar vw\right]$
\\
\hfill&\hfill&\hfill\\12&
$
-\sum_{i\in\Lambda^e_1}
\big[
 s^{e1\dag}_i\sigma^3s^{o1\   }_{i}
+s^{e1\dag}_i\sigma^3s^{o2\   }_{i-2\hat x_-}
-s^{e2\dag}_i\sigma^3s^{o1\   }_{i}
-s^{e2\dag}_i\sigma^3s^{o2\   }_{i}
$
\hfill&\hfill\\
\hfill&\hfill
$
+s^{e3\dag}_i\sigma^3s^{o3\   }_{i}
-s^{e3\dag}_i\sigma^3s^{o4\   }_{i}
+s^{e4\dag}_i\sigma^3s^{o3\   }_{i-2\hat x_-}
-s^{e4\dag}_i\sigma^3s^{o4\   }_{i}
+{\rm H.C.}
\big]
$
&
$-{\rm i}\int\ d^2x\left[\bar wv-\bar vw\right]$
\\
\hfill&\hfill&\hfill\\13&
$
-{\rm i}\sum_{i\in\Lambda^e_1}
\big[
 s^{e3\dag}_i\sigma^0s^{o2\   }_{i}
-s^{e3\dag}_i\sigma^0s^{o1\   }_{i}
+s^{e4\dag}_i\sigma^0s^{o2\   }_{i-2\hat x_-}
-s^{e4\dag}_i\sigma^0s^{o1\   }_{i}
$
\hfill&\hfill\\
\hfill&\hfill
$
-s^{e2\dag}_i\sigma^0s^{o3\   }_{i-2\hat x_+}
-s^{e2\dag}_i\sigma^0s^{o4\   }_{i-2\hat x_+}
+s^{e1\dag}_i\sigma^0s^{o3\   }_{i-2\hat x_+-2\hat x_-}
+s^{e1\dag}_i\sigma^0s^{o4\   }_{i-2\hat x_+}
-{\rm H.C.}
\big]
$
&
$+\int\ d^2x\left[\bar vz+\bar zv\right]$
\\
\hfill&\hfill&\hfill\\14&
$
-\sum_{i\in\Lambda^e_1}
\big[
 s^{e3\dag}_i\sigma^3s^{o2\   }_{i}
-s^{e3\dag}_i\sigma^3s^{o1\   }_{i}
+s^{e4\dag}_i\sigma^3s^{o2\   }_{i-2\hat x_-}
-s^{e4\dag}_i\sigma^3s^{o1\   }_{i}
$
\hfill&\hfill\\
\hfill&\hfill
$
-s^{e2\dag}_i\sigma^3s^{o3\   }_{i-2\hat x_+}
-s^{e2\dag}_i\sigma^3s^{o4\   }_{i-2\hat x_+}
+s^{e1\dag}_i\sigma^3s^{o3\   }_{i-2\hat x_+-2\hat x_-}
+s^{e1\dag}_i\sigma^3s^{o4\   }_{i-2\hat x_+}
+{\rm H.C.}
\big]
$
&
$-{\rm i}\int\ d^2x\left[\bar vz-\bar zv\right]$
\\
\hfill&\hfill&\hfill\\15&
$
-{{\rm i}\over2}\sum_{i\in\Lambda^e_1}
\big[
 s^{e1\dag}_i\sigma^0s^{e3\   }_i
+s^{e1\dag}_i\sigma^0s^{e4\   }_i
-s^{e2\dag}_i\sigma^0s^{e3\   }_i
-s^{e2\dag}_i\sigma^0s^{e4\   }_i
-(e\leftrightarrow o)
-{\rm H.C.}
\big]
$
&
$+\int\ d^2x\left[\bar vv-\bar zz\right]$
\\
\hfill&\hfill&\hfill\\
\end{tabular}
\label{table1}
\end{table}


\begin{references}
\bibitem{Mottinsulator} We understand under Mott insulator
an insulator with an odd number of electron per unit cell.
\bibitem{particle-hole} The identities
$
|0>_{s_i}=
\psi^{\dag}_{i2}|0>_{\psi_i}
$,
$
s^{\dag}_{i\uparrow  }|0>_{s_i}=
\psi^{\dag}_{i1}\psi^{\dag}_{i2}|0>_{\psi_i}
$,
$
s^{\dag}_{i\downarrow}|0>_{s_i}=
|0>_{\psi_i}
$,
and
$
s^{\dag}_{i\uparrow  }s^{\dag}_{i\downarrow}|0>_{s_i}=
\psi^{\dag}_{i1}|0>_{\psi_i}
$,
where $|0>_{s_i}$ is the spinon vacuum at site $i$, hold.
\bibitem{center} The center ${\rm C_{SU(2)}}$ of the group SU(2) is the set of
all matrices of SU(2) which commute with all other elements of SU(2).
It is made of two elements, the unit matrix $+\sigma^0$ and $-\sigma^0$.
As a group, the center is isomorphic to ${\cal Z}_2$:
$
{\rm C_{SU(2)}}=
\{
-\sigma^0,+\sigma^0
\}
\cong
\{
-1,+1
\}
=
{\cal Z}_2.
$
The adjoint representation maps the two elements of the center on the
3 by 3 unit matrix whereas the fundamental representation of SU(2) is
SU(2) itself.
\bibitem{PLforallWL} Any finite dimensional representation
of SU(2) can be constructed from the tensor product of the fundamental
representation. Physically, any source (color charged particle)
carrying the color charge of a higher dimensional representation than the
fundamental, can be thought as a composite of the fundamental sources
(e.g., hadrons are composite of quarks).
\bibitem{determinantfluctuations}
The usual justification for this approximation is that
the fluctuations of the determinant cost an energy of order $J_{2(1)}$,
which is not the lowest scale of energy due the softness of the fluctuations
of the SU(2) degrees of freedom. This approximation does not affect the
issue of confinement as shown in the
the discussion based on the $(\beta,\kappa,\infty)$ phase diagram
of section \ref{sec:Deconfinement}. However, this approximation
does not allow the effective continuum theory to be unstable towards
dimerization, an issue which is of particular importance for the one
dimensional antiferromagnet with competing nearest and
next-nearest-neighbor interactions.
\bibitem{dimerinstability} Since we are including determinant fluctuations
of the nearest-neighbor links $M$, our effective theory can,
in principle, detect an unstability of the mean-field theory towards
dimerization on nearest-neighbor links.
\bibitem{digressiononBZA}
The same argument applies to the BZA limit $J_2=0$ in which we expect
a ground state with N\' eel order, but there we do not
have an effective theory to study how the quantum fluctuations
affect the mean-field Ansatz.
\bibitem{dangerousirrelevance} One might wonder why the valence bond (dimer)
ordering results from the breaking of a continuous symmetry
in the continuum theory
when on the lattice it is only associated with the breaking of
discrete symmetries. This is because we have neglected (dangerous)
irrelevant operators in our gradient expansion
which modify the symmetries of the continuum theory.
\end{references}
\end{document}